\documentclass[10pt,conference]{IEEEtran}

\usepackage{amssymb}

\usepackage{amsmath,amsfonts}
\usepackage{algorithmic}
\usepackage{algorithm}
\usepackage{array}
\usepackage[caption=false,font=normalsize,labelfont=sf,textfont=sf]{subfig}
\usepackage{textcomp}
\usepackage{stfloats}
\usepackage{verbatim}
\usepackage{graphicx}
\usepackage{amssymb}
\usepackage{textcomp}
\usepackage[table]{xcolor}
\usepackage{tikz}
\usepackage{multicol}
\usepackage{wrapfig}
\usepackage{kantlipsum}
\usepackage{boldline}
\usepackage{float}
\usepackage{xspace}
\usepackage{booktabs, makecell, multirow, tabularx}
\usepackage[hyphens]{url}

\newcommand{\Henon}{\textit{Henon}\xspace}
\newcommand{\Lorenz}{\textit{Lorenz}\xspace}
\newcommand{\zstd}{\textit{zstd}\xspace}
\newcommand{\logistic}{\textit{Logistic}\xspace}

\definecolor{dkgreen}{rgb}{0,0.6,0}
\definecolor{gray}{rgb}{0.5,0.5,0.5}
\definecolor{mauve}{rgb}{0.58,0,0.82}

\newcommand*\circled[1]{\tikz[baseline=(char.base)]{
            \node[shape=circle,fill,inner sep=0.3pt] (char) {\textcolor{white}{#1}};}}

\newcommand*\circlenew[1]{\tikz[baseline=(char.base)]{
            \node[shape=circle,draw,inner sep=0.3pt] (char) {#1};}}    

\newcommand*\circlebig[1]{\tikz[baseline=(char.base)]{
            \node[shape=circle,draw,inner sep=0.3pt, minimum size= 5mm] (char) {#1};}}                

\newcommand{\name}{{\em SecOComp}\xspace}

\title{\name: A Fast and Secure Simultaneous Compression and Encryption Scheme}

\author{\IEEEauthorblockN{Nivedita Shrivastava}
\IEEEauthorblockA{Electrical Engineering Department\\
Indian Institute of Technology, Delhi, India\\
Email: nivedita.shrivastava@ee.iitd.ac.in}
\and
\IEEEauthorblockN{Smruti Ranjan Sarangi}
\IEEEauthorblockA{Electrical Engineering Department\\
Indian Institute of Technology, Delhi, India\\
Email: srsarangi@cse.iitd.ac.in}
}

\begin{document}
\maketitle
\thispagestyle{plain}
\pagestyle{plain}



\begin{abstract}
We live in a data-driven era that involves the
generation, collection and processing of a massive amount of
data. This data often contains valuable intellectual property and
sensitive user information that must be safeguarded. There is
a need to both encrypt and compress the data at line speed
and sometimes with added power constraints. The majority of
the currently available simultaneous compression and encryption
(SCE) schemes are tailored for a specific type of data such as
images for instance. This reduces their generic applicability. In
this paper, we tackle this issue and propose a generic, efficient,
and secure simultaneous compression and encryption scheme
where the data is simultaneously encrypted using chaotic maps
and compressed using a fast lossless compression algorithm.
We claim that employing multiple chaotic maps and a lossless
compression method can help us create not only an efficient
encryption scheme but also compress the data efficiently in a
hardware-friendly manner. We avoid all the known pitfalls of
chaos theory based encryption that have prevented its widespread
usage. Our algorithm passes all the NIST tests for nine different types of popular
datasets. The proposed implementation uses 1.51x less storage
as compared to the nearest computing work.


\end{abstract}





\section{Introduction}
\label{sec:intro}
The amount of digital data stored around the globe reached one zettabyte in the year 2012.
This signified the beginning of a new era - the Zettabyte era. 
This data includes everything from healthcare databases to trained neural network (NN) models.
Recent studies have projected an annual growth rate of 22\% for the cloud data storage, which means that it will double in size every four years~\cite{news1}.
The International Data Corporation (IDC) forecasts that the world's data will surpass 175 zettabytes by 2025, which will be more than triple the amount of data held in 2020~\cite{news2}.
This data explosion is a result of the incorporation of intelligent agents that utilize machine learning and other forms of artificial intelligence to assess the rising amount of data generated by digital devices in our daily lives.
This data is required during the development of driver assistance~\cite{driver}, autonomous car technologies~\cite{autonomous}, medical devices~\cite{medical}, IoT devices such as sensors in our bodies, homes, factories, and cities~\cite{IoT}, high-resolution content for virtual and augmented reality~\cite{VR}, and 5G communication systems~\cite{5G}.

Even a trained neural network model with tens of layers and millions of nodes suffers from the problem of high data storage cost. The model size is increasing day by day.
 Recently proposed OpenAI's ChatGPT~\cite{chatgpt} is one of the largest and most powerful language processing models to date with 175 billion parameters. As the volume of data increases, it is imperative that we focus on the most cost-effective means of storing it, particularly in settings with resource constraints. This issue also bedevils cloud-based servers where throughput and latency considerations dominate.

 To add to our woes, securing these data poses an additional concern as generating or acquiring this data is a costly affair. For an instance, acquiring raw data for training a neural network model itself can cost upwards of \$100K USD~\cite{hpca}. Thereafter, the know-how for the model and training process might be worth millions of dollars. This makes the trained model parameters as well as the consolidated raw data an attractive target for adversaries. Data is power, and if it is obtained by the wrong people, it can be used as a lethal weapon. Considering the amount of data generated today, it is crucial that we concentrate on techniques that can concurrently compress and encrypt the data since this will result in a significant reduction in compression and encryption overheads.  

Sadly, the research area of simultaneously looking at compression and encryption of data has mostly focused on securing images. These approaches perform encryption using image-based characteristics, which limits their applicability to a broader domain.
Another difficulty is that these algorithms are built for small-size data; hence, they are unsuitable for large data sets. Moreover, performing encryption and compression separately leads to additional overheads. For an instance, encryption-then-compression (ETC) is an unacceptable scheme as encryption decreases data redundancy and increases randomness, which negatively impacts the compression efficiency.
Another option, compression-then-encryption (CTE), is a slow method due to the sequential nature of compression algorithms.
Moreover, the difficulty of cracking the CTE scheme is equivalent to the difficulty of breaking the encryption algorithm. 
 
All of these factors underscore the necessity for a simultaneous compression and encryption (SCE) solution that is efficient and is applicable to big data as well.
In this work, we propose a variant of \textit{chaos-based encryption-compression}, which offers robust performance and security guarantees while retaining a reasonable compression ratio. We utilize \textit{zstd}, a fast lossless compression technique introduced by Facebook in 2016 for real-time compression scenarios.

Despite the prominence of chaos theory in the last decade, its impact on security research has been minimal.
This is because of three factors:
\circlenew{1}
The chaos-based approaches rely on extensive real number arithmetic, making practical implementations onerous.
\circlenew{2}
The implementations were typically slow.
\circlenew{3} The schemes were cryptographically weak and were prone to several attacks~\cite{attack1,cryptanalysis1,attack3}. Many researchers examined existing chaotic maps and demonstrated that if the chaotic performance is subpar, the maps become less secure~\cite{attack1,attack2}. Wang et al.~\cite{attack2}, \cite{R4Logistic}, {\cite{R2Image}, and Xi et al.~\cite{attack1} pointed out flaws in the encryption schemes based on simple 1-D chaotic maps. All of these factors had a negative impact on the acceptance of choas-based encryption techniques. 

However, the past three years have witnessed a \textit{resurgence}~\cite{chaos1},\cite{chaos2},\cite{chaos3},\cite{chaos4},\cite{chaos5},\cite{chaos6},\cite{chaos7},\cite{chaos8}\cite{chaos9},\cite{chaos10},\cite{chaos11}, \cite{chaos12},\cite{helper}, especially in the hardware community primarily because of the emergence of new improved algorithms. 
More intricate high-dimensional chaotic maps have been proposed~\cite{chaos6, helper} in 2021, which can offer high levels of security guarantees. Using these newly proposed chaotic maps, several authors~\cite{multiple,double} have created more secure chaos-based encryption techniques that rely on multiple chaotic maps. Although these algorithms proved to be resistant to a variety of attacks, they had substantial performance overheads. We found that these techniques are not intended for handling a lot of data. For example, when used to encrypt other large datasets, the encryption scheme proposed by Wang et al.~\cite{chaos6} is very slow because of its highly sequential design (24$\times$ slower than our scheme \name).   
To solve this problem, we rely on the work by Zhang et al.~\cite{henonvhdl} and Lahcene et al.~\cite{lorentzvhdl} who proposed an optimized fixed-point \textit{Henon} and \textit{Lorenz} map based design that is optimized for FPGAs, albeit in a generic context. This solved our performance issues because such algorithms are parallel, realizable on hardware, and also use fixed-point arithmetic. 

We also took a thorough look at all the recent attacks that are proposed to break the chaos-based encryption algorithms~\cite{R1ModXor}, \cite{R2Image}, \cite{R3Image}. Our selection of a specific combination of chaotic maps was based on an extensive examination and analysis of all the popular and relevant attacks. \textit{We found that no attack has been developed for a chaos-based encryption system that incorporates 1D, 2D, and 3D chaotic maps. The majority of attacks are trivial and target very simple 1D chaotic maps like the Logistic map}~\cite{R4Logistic,R1ModXor}. There is a body of theoretical work now that has generalized the space of attacks on chaos-based methods. Some generic properties have been identified -- if an algorithm satisfies any of these properties then it is deemed to be vulnerable. One such work is proposed by Chen et al.~\cite{attack3}.
They argue that the conventional approach where a chaotic map is used to generate pseudorandom numbers and the ciphertext is a XOR of this sequence and the plaintext, does not provide bulletproof security. To address all of these issues, we use three different non-trivial, multi-dimensional maps that have till date not been broken, we do not exclusively rely on the XOR operation, we permute the data based on the key. {\bf We believe that this choice of design decisions steers clear of all known attacks (incl. foreseeable extrapolations) and does not satisfy any of the properties of algorithms that make it vulnerable (as identified in highly cited prior work)}.

Now, coming to the advantages of chaos-based encryption, we need to understand that it has distinct steps:
first perform a permutation and then perform substitutions~\cite{helper}. 
The permutation preserves the data statistics, which can be exploited for compression~\cite{helper}. Naturally, this makes chaos-based algorithms better candidates for SCE.

As of today, the performance and security guarantees of software implementations of CTE and SCE~\cite{helper} algorithms for images have been extensively analyzed. 
This paper proposes a novel chaos-based encryption-compression technique for big data. In addition, we also designed and implemented an efficient HW circuit that pre-computes and generates chaotic maps in the shadow of DRAM reads and compression engine operations, which leads to a much faster execution time. 

The \textbf{main contributions in this paper} are as follows:  \circlenew{1} Elaborate characterization of various chaotic maps and compression algorithms for NN workloads. \circlenew{2} Insights from the characterization section to find the right combination for designing a low-overhead and fast SCE algorithm to efficiently compress and encrypt the weights. \circlenew{3} A novel permutation scheme using the \textit{MergeShuffle} algorithm. \circlenew{4} A novel highly-parallel HW design of the SCE engine, where we overlap the shuffling process with the compression operation to make the design faster. We also pre-compute the chaotic maps parallely in the shadow of DRAM writes and the compression process. \circlenew{5} A rigorous experimental and security analysis.

We have considered nine different types of datasets: neural networks, images, Kaggle datasets (Chatbots, medical data, etc). We shall extensively analyze pre-trained and quantized NN model data. The additional results of other datasets, which are quite similar, are also summarized at the end of Evaluation section. Our algorithm has been publicly accessible since Feb, 2023, and even after popularizing the algorithm, nobody has been able to propose a successful attack.

The text is divided into 8 sections.
$\S$\ref{sec:back} provides the background. $\S$\ref{sec:attack} outlines an attack on a CTE scheme. $\S$\ref{sec:char} provides an characterization of compression and encryption schemes. $\S$\ref{sec:hw} presents the proposed hardware design, $\S$\ref{sec:eval} presents the performance results, $\S$\ref{sec:security} presents the security analysis and $\S$\ref{sec:RW} presents the related work. We finally conclude in $\S$\ref{sec:conc}. 
\section{Background}
\label{sec:back}
In this section, we present the relevant background.
\subsection{Chaotic Maps}
A chaotic map~\cite{logistic} is a discrete/continuous multi-dimensional map (multi-dimensional matrix) with a very high dependence on initial conditions. Some excellent properties of chaotic maps, such as sensitivity to initial conditions, pseudo-randomness, ergodicity, and a large system parameter space, make them a popular choice for designing an encryption algorithm. 

The idea that chaotic systems are sensitive to initial conditions indicates that minuscule variations in initial conditions result in significant long-term variations. This is one of the defining features of chaos. We explain the basics of chaotic maps using a \logistic map. We represent the \logistic map using the equation $x_{i+1}=\mu x_i(1-x_i)$, as shown in Table \ref{tab:mapsEqu}.
The new value, denoted by $x_{i+1}$, is determined by the previous value, $x_i$, as well as the control parameter $\mu$.
A slight fluctuation in $\mu$ results in long-term changes in the map.
This behaviour is illustrated in Figure \ref{fig:bifur}, which shows the bifurcation diagram for the \logistic map that demonstrates how the values of one of the variables of a chaotic map change as the control parameter is varied.

The diagram indicates that fixed values are attained only when $\mu\leq3$.
When $\mu=$ 3.1, a bifurcation is visible on the map.
The computed value oscillates. When $3.4 \leq \mu \leq 3.5$, the expected value oscillates between four distinct values as shown in Figure \ref{fig:bifur}. As the value of $\mu$ increases, the set of values continue to multiply until chaotic behavior sets in (all possible values within the range subject to the real number resolution). We can clearly see that chaotic behavior sets in when $\mu \geq$ 3.6.


\begin{figure}
    \centering
    \includegraphics[scale=0.11]{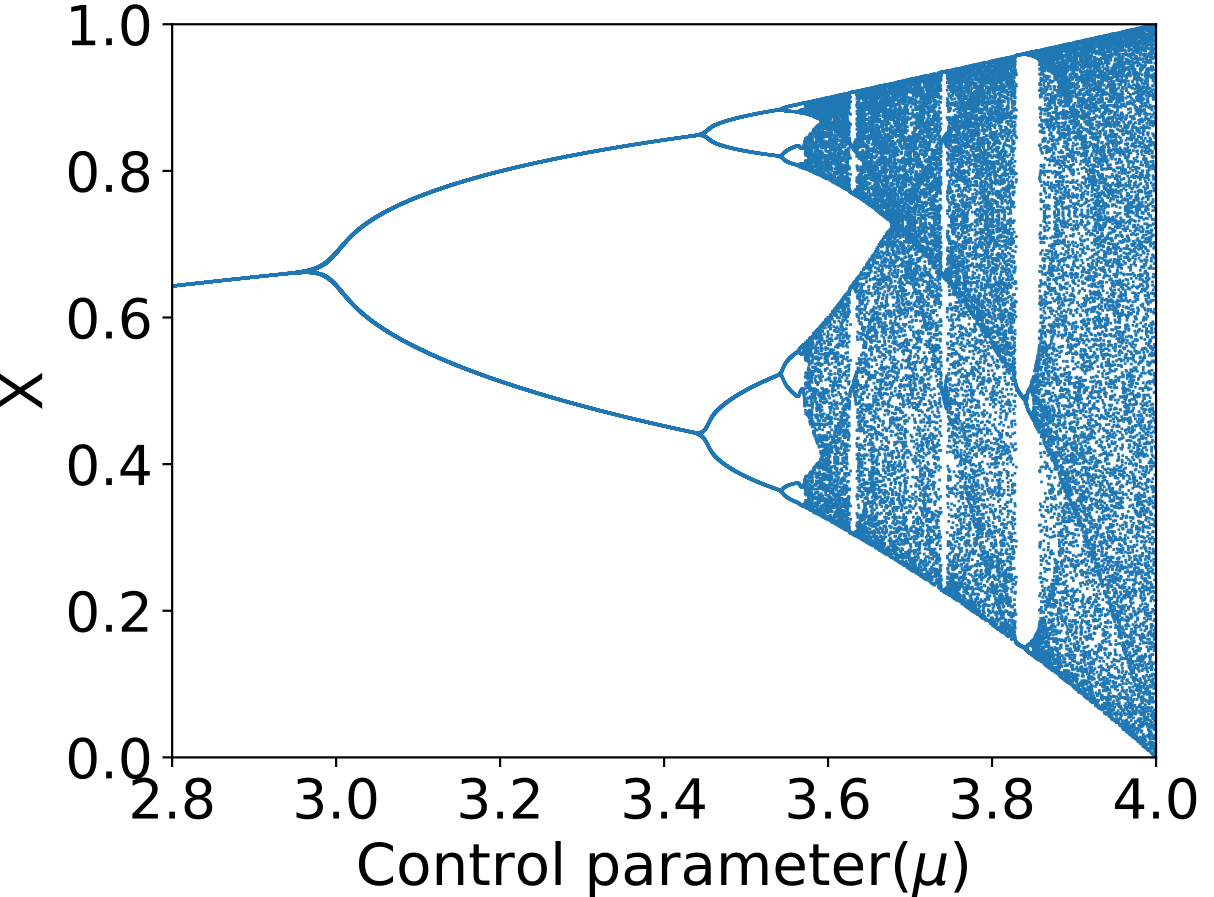}
    \caption{The Bifurcation diagram for the \logistic map}
    \label{fig:bifur}
\end{figure}

\begin{figure}
    \centering
    \includegraphics[scale=0.27]{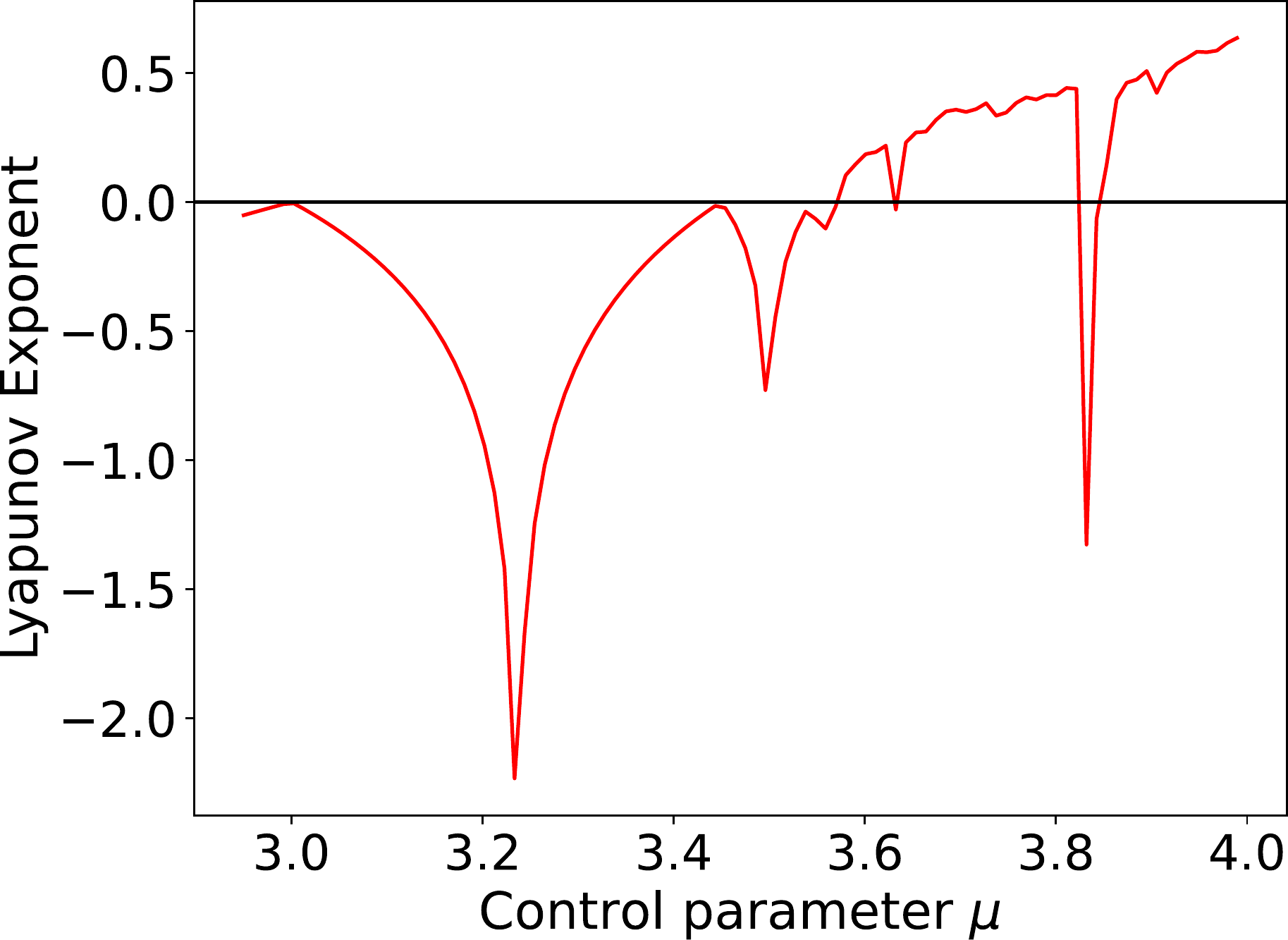}
    \caption{Variation of the Lyapunov exponent with the control parameter for the \logistic map}
    \label{fig:LE}
\end{figure}

The chaotic maps that we consider are shown in Table \ref{tab:mapsEqu}.

\begin{table}[h]
\footnotesize
    \centering
    \begin{tabular}{|p{15mm}|l|}
    \hline
    \rowcolor{blue!10}
    \textbf{Chaotic Map} & \textbf{Equations}  \\
    \hline
    Henon~\cite{henon} &  $x_{i+1}=1+y_i-a{x_i}^2$; $y_{i+1}=bx_i$\\
    Tent~\cite{tent} &  $x_{i+1}$ =  
    $\mu x_i$ if $x_i<1/2$;
    $x_{i+1}$ = $\mu (1-x_i)$ if $x_i>1/2$     \\
    Logistic~\cite{logistic} & $x_{i+1}=\mu x_i(1-x_i)$\\
    Lorenz~\cite{lorentzvhdl} & $x_{i+1}=\sigma (y_i-x_i)$; $y_{i+1}=x_i(\rho-z_i)-y_i$;\\ 
    & $z_{i+1}=x_iy_i-\beta z_i$\\
    Chirikov~\cite{chirikov} & $x_{i+1}=x_i+ksin(y_i)$ ; $y_{i+1}=y_i+x_{i+1}$\\
    \hline
    \end{tabular}
    \caption{Chaotic maps used in the paper. (x, y, z) represents map values while other variables represent control parameters.}
    \label{tab:mapsEqu}
\end{table}

\subsubsection{Lyapunov Exponent (LE)}
It serves as a measure of the sensitivity and predictability of the dynamic system with respect to changes in its initial conditions. A zero LE indicates that a system has reached the steady state. However, a negative LE indicates that a system has asymptotic stability. Any system with a positive LE is a chaotic system. 
 For a discrete time chaotic map $x_{n+1}=f(x_{n})$, LE ($\lambda$) is calculated as  $     \lim_{N\to \infty} \frac{1}{N} \sum ln 	\left|  \frac{dx_{n+1}}{dx_n} \right| $

Figure \ref{fig:LE} illustrates how the LE varies with the control parameter for the \logistic map. We can see that the LE approaches a positive value as soon as chaos sets in ($\mu \geq 3.6$).

While encrypting using chaotic maps, the {\bf secret key} is a combination of the initial value of $x$ and the value of control parameters (within a pre-defined range).
\vspace{-3mm}
\subsection{Compression Techniques}
 Modern compression algorithms rely on a combination of traditional compression methods such as the following.\\
\textbf{Huffman encoding} translates the data into unique binary codes. The binary code for the character with the highest frequency is the shortest, while the one with the lowest frequency is the longest. \\
\textbf{LZ-77 Encoding} uses the sliding window concept to reference previously read data/phrases. If a word/phrase is repeated, the previous occurrence is marked with a reference point (dictionary-based scheme). \\
\textbf{Run-Length Encoding} Repetitive symbols in the data are replaced with a \textit{special character}, a \textit{symbol}, and the \textit{binary count} of the matching repetitive string.
Each repeating string is represented by three fields: special character, symbol, and count. 
Table~\ref{tab:compress} presents a short description of the modern compression algorithms.
\begin{table}[!h]
\footnotesize
    \centering
    \begin{tabular}{|p{1cm}|p{0.68cm}|p{5cm}|}
    \hline
    \rowcolor{blue!10}
    \textbf{Algo.} & \textbf{Version} & \textbf{Based on} \\
    \hline
        deflate~\cite{deflate} & 1.10 & LZ77, Huffman coding\\
        bzip2~\cite{bzip2} & 1.0.8 & Burrows-Wheeler algo., Huffman Coding\\
        lzma~\cite{lzma} & 5.2.5 & Sliding window
technique of LZ77  \\
        zstd~\cite{zstd} & 1.4.5 & LZ77, Huffman or run-length encoding\\
         \hline
    \end{tabular}
    \caption{Various lossless compression libraries used in the work}
    \label{tab:compress}
\end{table}

\vspace{-5mm}

\subsection{NIST Test}
The NIST~\cite{nist} test suite is the gold standard for testing the security of encryption algorithms. This test is widely used to test the efficiency of various encryption ciphers including AES~\cite{nistAes}. The NIST test suite comes with 15 tests~\cite{nist} as shown in Table~\ref{tab:nistTest}. The details of the tests have been explained in detail by Marton et al.~\cite{nistRange}. Each test assesses whether the ciphertext appears to be random and whether it yields any statistical information about the corresponding plaintext, data patterns and the key. The test generates a \textit{p}-value that indicates whether the ciphertext would be considered as random or not.
It is regarded as random if the \textit{p}-value exceeds a pre-specified significance level $\alpha$ (typically $0.01$).

\begin{table}[h]
\footnotesize
    \centering
    \begin{tabular}{|l|l||l|l|}
    \rowcolor{blue!10}
    \hline
     \textbf{Test Name} & \textbf{Rep.} &  \textbf{Test Name} & \textbf{Rep.} \\
     \hline
      Frequency  &  T0 & Overlapping & T8\\
      BlockFrequency   & T1 & Universal & T9 \\
      CumulativeSums & T2 & ApproximateEntropy & T10\\
      Runs & T3 & Random Excursion & T11\\
      LongestRun & T4 & Random Excursion Variant & T12\\
      Rank & T5 & Serial & T13\\
      FFT & T6 & LinearComplexity & T14\\
      Nonoverlapping & T7 & & \\
      \hline
    \end{tabular}
    \caption{NIST Test Suite}
    \label{tab:nistTest}
\end{table}
\vspace{-5mm}

\section{An Attack on CTE}
\label{sec:attack}
\noindent \textbf{Notation-} Let the victim's data $P$ (of size $S_P$) be fed into the CTE engine.
In a typical CTE algorithm, the compression engine compresses $P$ to produce intermediate compressed data $P_C$, which is then sent to the AES engine to produce the final crypto-compressed data $C$ of size $S_C$. 

\noindent \textbf{Threat model-}
The victim and attacker processes are executing on a system where the last level cache is shared.
The attacker does not have access to $P$, $P_C$ or the encryption key ($K$).
She has access to $S_P$ and $S_C$ by analyzing the memory traffic.
She can also access $C$.
The attacker knows the type of encryption scheme, but not the compression scheme.
The objective of the attack is to recover $K$ and $P$. This is the standard threat model.

\noindent \textbf{Attack-}
As the compressed data is encrypted, the attacker needs to first break the encryption.
She can deduce the compression ratio by comparing the initial and final output sizes and the secret key using a timing-based side channel attack. 
Let us elaborate.

AES-128 does 10 rounds of encryption; rounds 1-9 encrypt the data using T-tables $T_0,T_1,T_2,T_3$.
The final round is special as it uses a distinct T-table $T_4$; this provides the added benefit of reduced noise. The final round ciphertext is computed as : 
$C_i=k_i \oplus T_4[j]$, where $i$ represents the byte of the subkey and $j$ represents the T-table index. An attacker will monitor T-table lookups (using prime-probe or similar attacks) and will guess the key byte $k_i$ as $z$. If the guess is correct then the equation: $j= T_4^{-1}(C_i \oplus z)$ will be satisfied. In this way an adversary will be able to successfully retrieve the subkey similar to the previous works~\cite{finalRound}.

Our experiments show that the $inception$ net model of size 25 MB can be compressed to 8 MB which requires 500K AES-128 encryptions, which is enough to successfully mount a side-channel attack.
Even the smallest NN, $shufflenet$, requires 53k encryptions, which is sufficient for mounting an attack (as per prior work).

After obtaining the subkey, an attacker can reverse-engineer the original key(refer to~\cite{subkey}). Since she has both the key and the ciphertext, retrieving $P_C$ is simple. 
The attacker now has access to the compression ratio and $P_C$.
She can obtain $P$ by determining the type of compression algorithm.
This is a solved problem~\cite{compressNN}; most solutions use NNs to deduce the algorithm out of a known set of algorithms. 

In contrast to this, our scheme provides multi-level security by adding an additional permutation layer before the compression stage and a substitution layer after compression. Additionally, \textit{we generate the chaotic maps simultaneously, which distorts the micro-architectural hardware signature of the map generation process.} This reduces the chances of side-channel attacks because of the added noise~\cite{esl}.

\section{Characterization of Compression Algorithms and Chaotic Maps for CNN Models}
\label{sec:char}

We characterize the behaviour of different compression algorithms using the setup mentioned in Table \ref{tab:config}. We executed different NN benchmarks, which are pre-trained and quantized for the ImageNet dataset. We used quantized models from the PyTorch Torchvision framework (see Table ~\ref{tab:bm}). The benchmarks are chosen based on the availability of the pre-trained and quantized model; this is to ensure that our evaluation is realistic. We would like to add that \name will work efficiently for models of any size.

\begin{table}[h]
    \footnotesize
    \centering
    \begin{tabular}{|p{35mm}|p{38mm}|}
    \hline
   \rowcolor{blue!10}
    \multicolumn{2}{|c|}{\textbf{Hardware Settings}}  \\
    \hline
    Intel Core i5-8300H CPU, 2.3 GHz & DRAM: 8 GB \\
    \hline
    \#CPUs: 8 Cores & L1:128 KB, L2:1 MB, L3:8 MB \\
    \hline
    \rowcolor{blue!10}
    \multicolumn{2}{|c|}{\textbf{Software Settings}} 
    \\
    \hline
    Linux kernel: 5.15 & Operating System: Ubuntu 20.04 \\
    \hline 
    Python version: 3.8 & Pytorch version: 1.11 \\
    \hline
    GCC: 9.4 & Boost version: 1.71 \\
    \hline
    \hline
    \end{tabular}
    \caption{System Configuration for Software Characterization}
    \label{tab:config}
\end{table}

\subsection{Characterization of Compression Algorithms}
We examine the compression ratio and the compression time for the following lossless compression algorithms: \textit{zstd}, \textit{deflate}, \textit{zlib}, \textit{bzip2}, and \textit{lzma}.
As the algorithms are lossless, the accuracy of the NN models will not be affected; hence,  accuracy analysis is unnecessary. We estimate the \textbf{compression ratio} which is the ratio between the size of the original model and the compressed model, as well as the compression time, as shown in Figures \ref{fig:compRatio} and \ref{fig:compTime}. 

We found that the \textit{bzip2} algorithm gives the best compression ratio while \textit{zstd} gives the best compression time. We observe that the compression time is proportional to the model size. Although \textit{zstd} does not provide the best compression ratio, but the ratio is still comparable with the other algorithms. Additionally, \textit{zstd} is 7.49$\times$  faster than \textit{bzip2}. Moreover, \textit{zstd} is only 1.27$\times$ worse than \textit{bzip2} with respect to the compression ratio. Thus, we decided to use \textit{zstd} for compressing NNs owing to its superior performance.



\begin{table}[!h]
\footnotesize
    \centering
    \begin{tabular}{|l|p{18mm}||p{14mm}|p{18mm}|}
    \hline
    \rowcolor{blue!10}
    \textbf{Benchmark} & \textbf{Model Size (MB)} &\textbf{Benchmark} & \textbf{Model Size (MB)} \\
    \hline
        googlenet~\cite{survey} & 6.89 & inception~\cite{survey} & 24.36\\
        \hline
        mobilenet~\cite{survey} & 3.63 &
        resnet~\cite{survey} & 11.84  \\
        \cline{1-2}
        shufflenet~\cite{survey} & 2.56 & &\\
         \hline
    \end{tabular}
    \caption{A Description of neural network benchmarks}
    \label{tab:bm}
\end{table}

\begin{figure}[!h]
    \centering
    \includegraphics[scale=0.27]{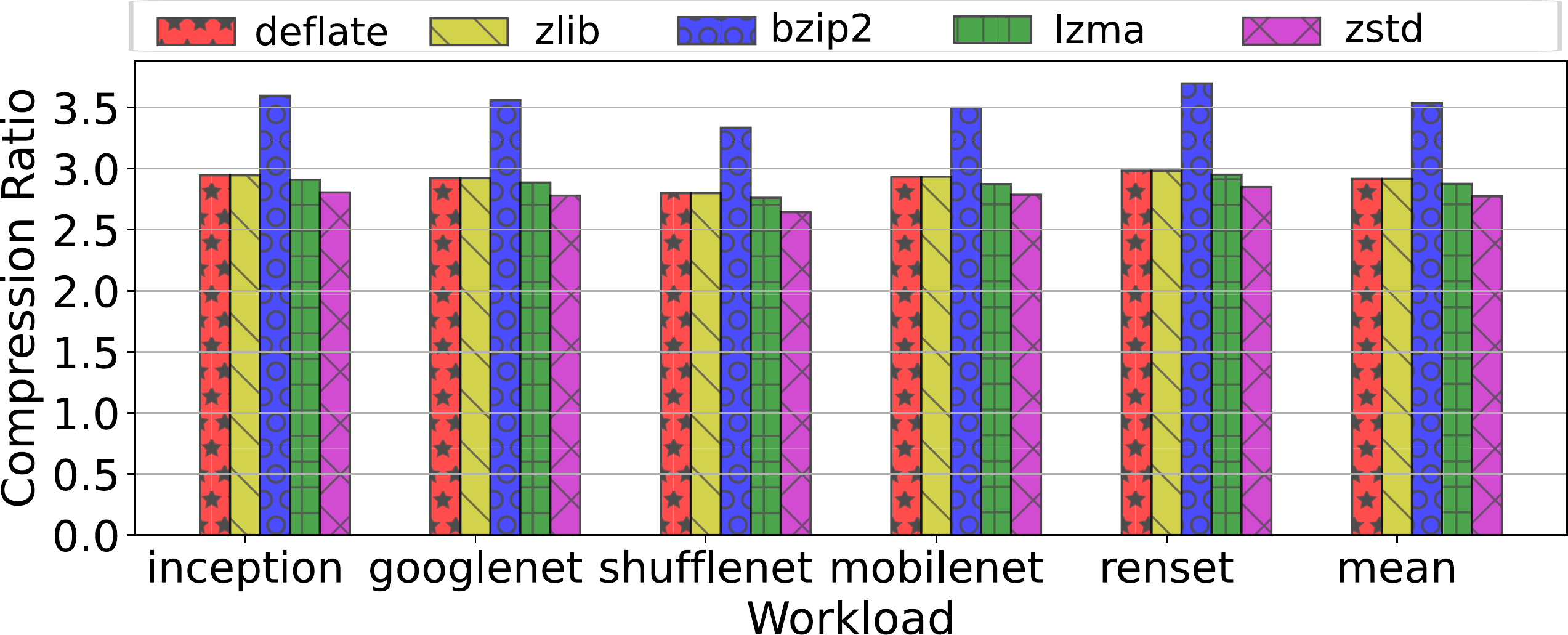}
    \caption{Characterization of the compression algorithms (on the basis of the compression ratio)}
    \label{fig:compRatio}
\end{figure}

\begin{figure}[!h]
    \centering
    \includegraphics[scale=0.27]{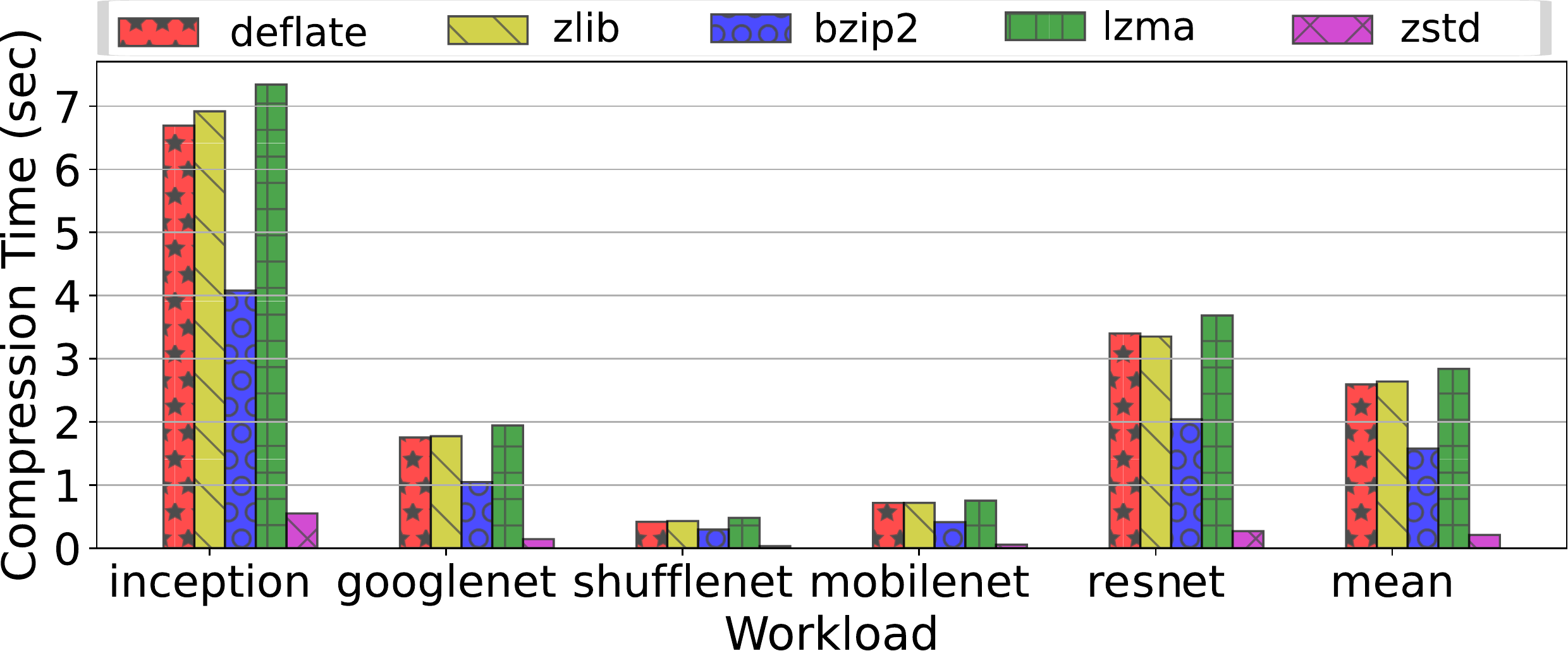}
    \caption{Characterization of the compression algorithms (on the basis of the compression time)}
    \label{fig:compTime}
\end{figure}

\subsection{Characterization of Chaotic Maps}
The objective of this experiment is to determine and select the chaotic maps that have the lowest correlation coefficients and map generation times. The correlation coefficient (CC) measures the linear relationship between the two datasets. We compute the \textit{Pearson's correlation coefficient} (CC) as
   $\frac{cov (X,Y)}{\sigma_X \times \sigma_Y}$
 Here, $cov(X,Y)$ represents the covariance between the variables $X$ and $Y$, while $\sigma$ represents the standard deviation. 
A high correlation coefficient represents a strong relationship between the variables and vice-versa. 

We selected the most popular chaotic maps for characterization as shown in Table \ref{tab:maps}. We performed experiments to estimate the constant values, which would result in chaotic behavior by estimating the Lyapunov Exponent (LE) as presented in Table \ref{tab:maps}. 

Thereafter, we used the constants to generate the chaotic maps and performed a XOR operation between the maps and the weights. We present the results in Figures \ref{fig:mapCC} and \ref{fig:mapTime}. We found that the \textit{Chirikov} map is one of the slowest maps due to repeatedly computing the sine function.

\begin{table}[h]
\footnotesize
    \centering
    \begin{tabular}{|l|l|l|}
    \hline
    \rowcolor{blue!10}
    \textbf{Benchmark} & \textbf{Constants} & \textbf{LE} \\
    \hline
    Henon~\cite{henon} (2D) & a=1.4; b=0.3 & 0.61   \\
    \hline
    Tent~\cite{tent} (1D)  & $\mu$=1.98  & 0.68 \\
         \hline
    Logistic~\cite{logistic} (1D) & $\mu$=3.98 & 0.63 \\
    \hline
    Lorenz~\cite{lorentzvhdl} (3D) & $\sigma$=10, $\rho$=28, $\beta$=-2.67 & 0.92 \\
    \hline
    Chirikov~\cite{chirikov} (2D) & k=10  & 0.85 \\
    \hline
    \end{tabular}
    \caption{Control parameters of chaotic maps }
    \label{tab:maps}
\end{table}

\begin{figure}[!h]
    \centering
    \includegraphics[scale=0.27]{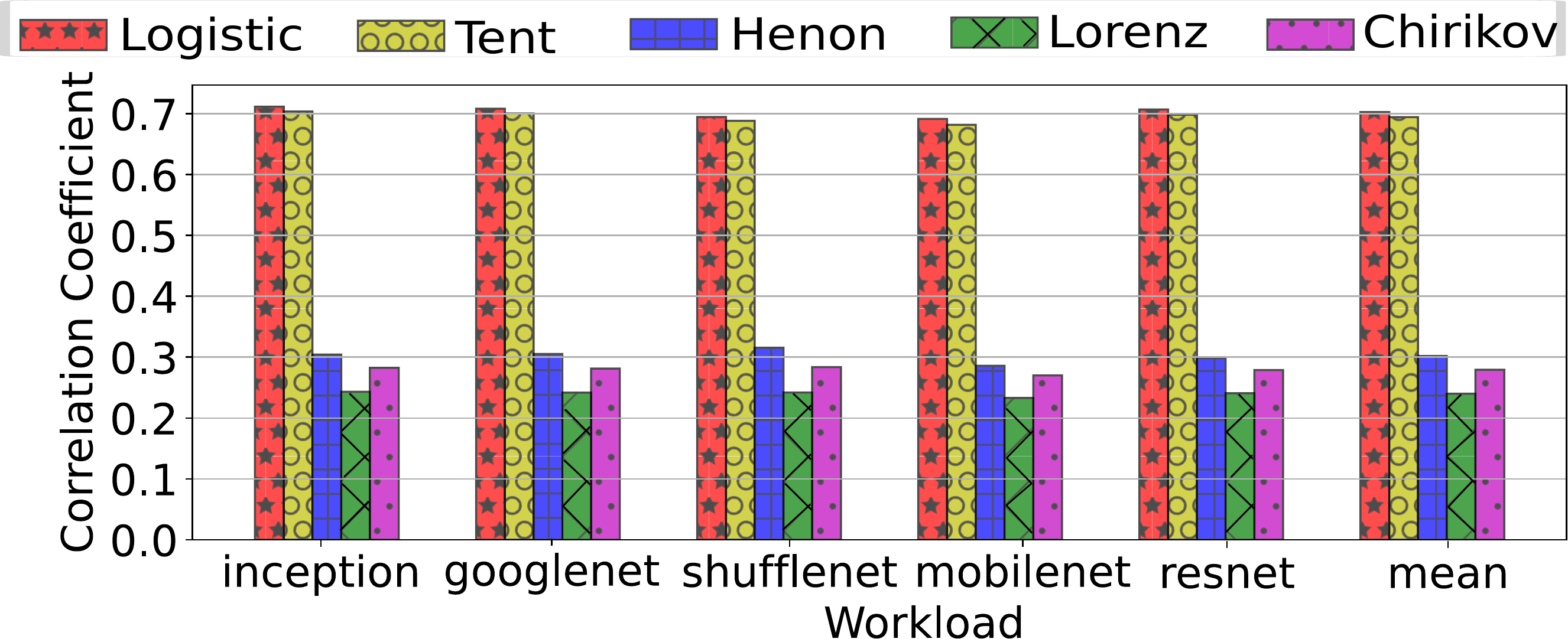}
    \caption{Characterization of the chaotic maps (on the basis of the correlation coefficient)}
    \label{fig:mapCC}
\end{figure}

\begin{figure}[!h]
    \centering
    \includegraphics[scale=0.27]{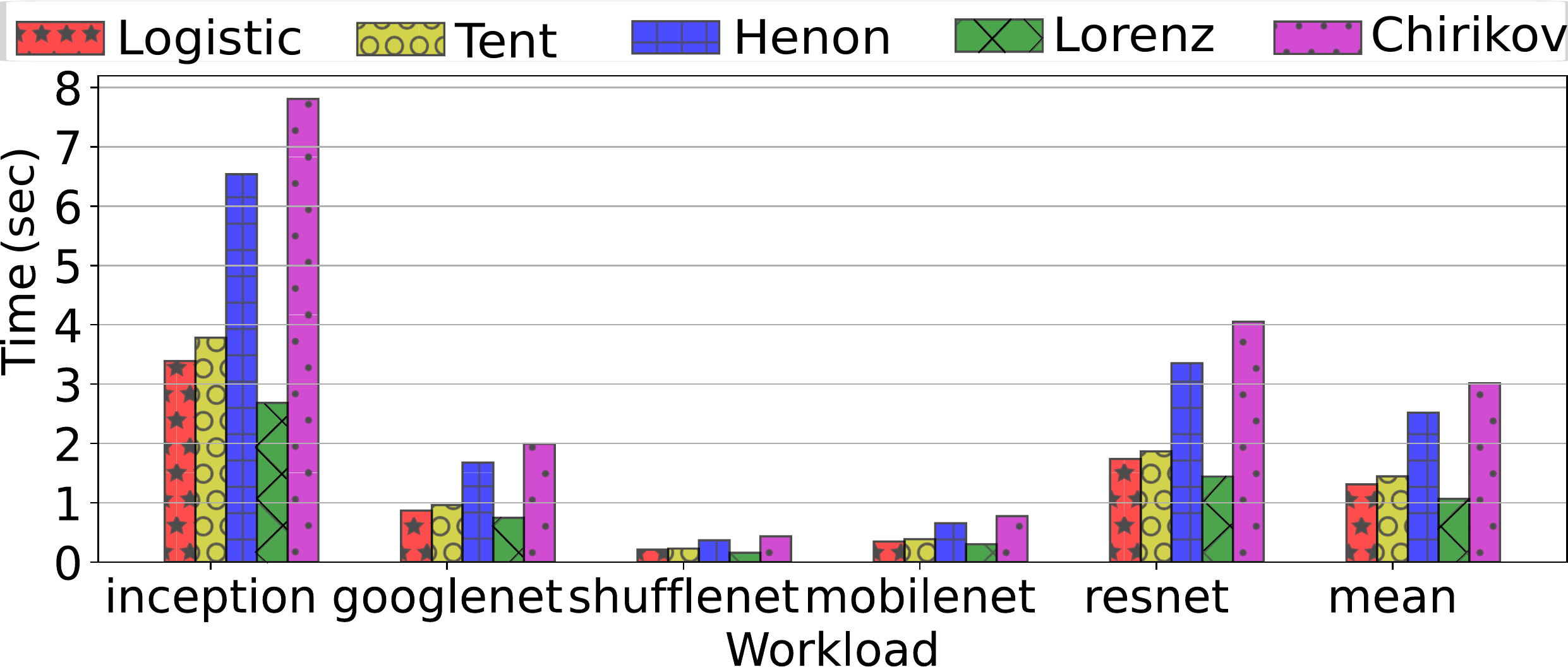}
    \caption{Characterization of the chaotic maps (on the basis of the map generation time)}
    \label{fig:mapTime}
\end{figure}

We choose three maps based on the results: \textit{Logistic}, \Lorenz, and \Henon. The \textit{Logistic} and \Lorenz maps have the fastest map generation times. The \Henon map takes longer to generate but its correlation coefficient is the second lowest (after \Lorenz). We did not choose \textit{Chirikov} because of its slow generation time and \textit{Tent} because of its low performance and high correlation coefficient.

{\bf Conclusion: The most appropriate choices are the {\em Logistic}, {\em Lorenz}, and {\em Henon} maps.
Among compression algorithms, \zstd is the fastest with a good compression ratio. }


\begin{figure}
    \centering
    \includegraphics[scale=1.4]{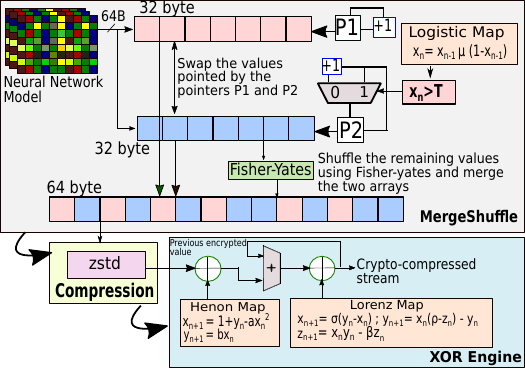}
    \caption{Overall proposed scheme}
    \label{fig:scheme}
\end{figure}

\begin{algorithm}
\caption{Permutation using chaotic maps}\label{alg:perm}
 \textbf{Input:} Intermediate data array $B={B_1,B_2....B_{64}}$; secret Threshold $T$ \\
 \textbf{Output:} Permutated array $B$
 
\begin{algorithmic}
\vspace{1.5mm}
\STATE \textbf{Initialize} $K[0] = k_0$; $ srt = 1$ ; $ mid = 32$ ; $ end = 64$\\
\noindent \nonumber{\textbackslash * Keep on iterating until one of the arrays is shuffled \textbackslash * }

\WHILE {$(strt \neq 32\ \&\ mid \neq 64)$}
  
  \STATE $L1 = LogisticMap()$
  \IF{$L1 > T$}
      \STATE $switch\ ( B [srt], B [mid] )$
\STATE  \noindent \nonumber{\textbackslash * Switch the values stored in the locations pointed by the srt and mid * }
      \STATE $mid \leftarrow mid+1$
   
   \ENDIF
   \STATE $strt \leftarrow strt+1$
   
\ENDWHILE\\
\vspace{1.5mm}
\noindent \nonumber{\textbackslash * Merge the remaining values in the undepleted array using the Fisher-Yates algorithm *\textbackslash }

  \WHILE{$mid \neq 64$}
    \STATE $pos = LogisticMap(\ ) \% (mid, 64)$
    \STATE $switch\ (B[mid], B[pos])$
    \STATE $mid \leftarrow mid+1$
  \ENDWHILE
\STATE \textbf{Return} B  
\end{algorithmic}
\end{algorithm}

\section{Design Implementation}

\label{sec:hw}

\subsection{Overview}
The \name hardware (SCE engine) consists of an permutation engine, a compression engine, a XOR-based substitution engine, and three compact hardware circuits that correspond to the chaotic map generators (see Figure \ref{fig:design}). 

We used the \textit{zstd} compression engine from \cite{zstdhardware} and combined it with our proposed hardware, which simultaneously pre-computes chaotic maps while DRAM reads and compression engine operations are going on. Let us elaborate. Please keep referring to Figure~\ref{fig:scheme} and \ref{fig:design}.

\subsection{Data Pre-processing and Quantization: Step (1)}
\circlebig{1} The service provider/owner of the model trains an NN model and quantizes its weights and biases. It sends the appropriately quantized values to the SCE engine for further compression and encryption. The quantization process is performed by the model owner before the data is written to memory. We assume that the hardware is aware of the data storage pattern of the NN model (standard assumption). 

\begin{figure}[h]
    \centering
    \includegraphics[scale=0.20]{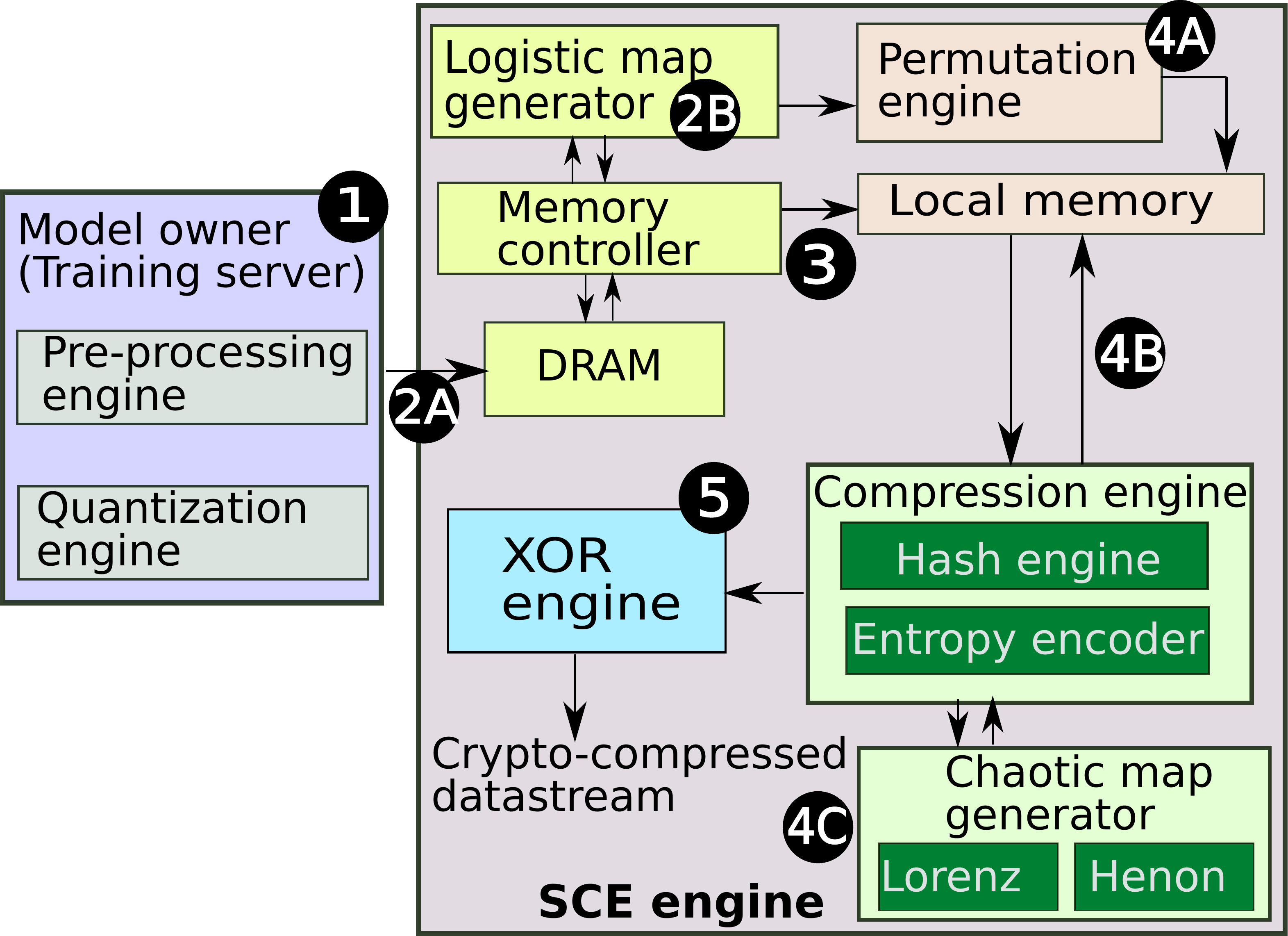}
    \caption{A high-level design of the scheme}
    \label{fig:design}
\end{figure}

\subsection{Data Read and Shuffling: Steps (2A), (2B), (3), and (4A)}

\circlebig{2A} The controller reads the weights from the DRAM and \circlebig{2B} simultaneously generates a \textit{Logistic} map. The aim of doing this is to permutate the values read from DRAM according to the values stored in the \textit{Logistic} map. We achieve three major benefits from this scheme: First, the map generation delay is subsumed in the shadow of DRAM reads. Second, permutation has little impact on the statistical features of data; this keeps the compression ratio intact. Third, due to the simultaneous map generation and DRAM reads, the micro-architectural signatures get distorted; this reduces side-channel leakage. 

\circlebig{3} We store this data in a local memory for subsequent computations.
Before the compression engine can compress this `ordered data', it is shuffled (permuted). For permutation, we use the \textit{MergeShuffle}~\cite{mergeshuffle} algorithm. The algorithm is shown in Algorithm \ref{alg:perm}.
We divide a 64-byte block into two equal 32-byte chunks. We associate two pointers (starting at 0) with these two chunks. 
The \textit{Logistic} map acts as a random number generator -- the $i^{th}$ Boolean random number is generated by comparing the $i^{th}$ byte of the \textit{Logistic} map with a secret threshold, $T$. If it is not equal, we swap the values indexed by the pointers and increment the pointers, otherwise we just increment one pointer without swapping. Ultimately, when one pointer reaches the end of a chunk, the remaining elements of the other chunk are permuted using the \textit{Fisher-Yates} algorithm~\cite{mergeshuffle}.

\circlebig{4A} This procedure is repeated for all subsequent DRAM reads.
 The procedure of shuffling is straightforward and does not appear in the critical path of the system because it completely overlaps with the process of compression.
Furthermore, the permutation engine operates at a higher frequency as compared to the compression engine, and while the compression of the current data bytes is taking place, the next chunk of data is fetched and permuted.
 Our tests show that this solution is necessary for creating a robust encryption algorithm and it also helps us maintain a good compression ratio.

\subsection{Compression Engine: Step (4B)}
\circlebig{4B} The compression module reads the shuffled weight values from the local memory. The module comprises two main stages: a compression stage (performed using hash engines) and an entropy encoding stage. The readers can refer to \cite{zstdhardware} for a detailed architecture of the compression engine.
In this work, we use a single compression unit. Our design can support many units as well. 

\subsection{Substitution Engine: Steps (4C) and (5)}
We use a combination of two chaotic maps for substitution: the 2-D \textit{Henon} map and the 3-D \textit{Lorenz} map. Experiments demonstrated that a single map is insufficient to encrypt the weights, as the weights encrypted using a single map fail to provide sufficient encryption. This motivated us to use a combination of chaotic maps (similar to previous work~\cite{helper}).
We choose the \textit{Logistic} map for permutation prior to compression because it is very fast and preserves important statistics about the data (given its high correlation coefficient).
On the other hand, \Lorenz and \Henon are two strong maps with a low correlation coefficient; their combination allows us to create a strong substitution engine. We are free to use the strong maps since the substitution is performed after the compression and the maps will have no impact on the compression ratio.

The encrypted initial values and the control parameters for the maps' generation are read from the DRAM. The chaotic map generators will continue to generate the values until they receive an interrupt from the compression engine to either pause or finish the value generation process. 


\circlebig{4C} As soon as the compression begins, a {\em start} signal is transmitted to both the map generators -- \Henon and \Lorenz.
The \textbf{benefit} of this strategy is that the map generation process overlaps with the  compression operation. A faster clock running at a 1 GHz frequency is used to generate the maps.

The maps generate 32-bit fixed-point values that are converted to 8-bit values as the weights are of size 1 byte (same as the previous work~\cite{helper}). 
Thus, 8 LSB bits are extracted from the 32-bit values and are then used to encrypt the weights. \circlebig{5} The encryption is performed using XOR gates and adders. The compressed information is XORed with the \Henon map. Following this, the generated value is added to the previous crypto-compressed value, and the result is again XORed with the \Lorenz map. This operation generates the next crypto-compressed value. 

\vspace{-2mm}
\subsection{Simultaneous Decompression and Decryption Unit}
The simultaneous decompression and decryption unit is simply a inverse of the SCE engine.
We re-generate the same chaotic maps (\Lorenz and \Henon) using the same initial values and control parameters; these maps are XORed with the crypto-compressed datastream.

Then, decompression is performed to generate the de-compressed data.
In the end, we re-generate the \textit{Logistic} map using the same initial conditions and the control parameters in order to de-shuffle the de-compressed data. This procedure will produce the actual data. Given that we used fixed-point arithmetic, this process is deterministic and portable across ALU designs.
\section{Evaluation}
\label{sec:eval}
In this section, we perform a thorough performance evaluation of the proposed scheme and compare with a few recent proposals. 
We show all the evaluated configurations in Table \ref{tab:configurations}. E1 is an ETC scheme with \zstd and AES (Cipher-Block Chaining (CBC) mode). C1 is a CTE scheme that is the reverse of E1. C2~\cite{chaos6} uses a chaotic encryption-based CTE scheme and \zstd for compression. S1 is the state-of-the-art in the SCE community proposed by Ahmad et al.~\cite{helper}. \name is our implementation, which has four specialized configurations of the form $Sec\_XYZ$, where $XYZ$ is the compression algorithm. Note that $zstd$ is the default and is used with \name.

\vspace{-3mm}
\subsection{Comparative Analysis (Software)}
\subsubsection{A comparison with S1 and C2 (related work)}
We used the setup shown in Table \ref{tab:config}. 
 We observe a $70\%$ improvement in the execution time and a $1.28\times$ better compression ratio using \name as compared to S1~\cite{helper} (see Figures \ref{fig:comp1} and \ref{fig:comp2}). The main reasons behind the performance deterioration in S1 are as follows \circled{1} In order to perform the permutation, they need to sequentially iterate through the entire dataset two times. This results in a significant performance overhead, particularly, if the dataset is large.
 However, in \name, only one iteration is required to perform the permutation. 

To improve the speed of Huffman coding, the authors in S1 use the \textit{Chinese Remainder Theorem (CRT)} to transform adjacent data elements into a unique CRT solution. 
\circled{2} The authors pre-compute the unique solutions of the CRT for two 8-bit values and store the resultant 16-bit solution in a look-up table (LUT). The size of the LUT is estimated to be $ 2^{8} \times 2^8 \times 16$ bits $= 2^{20}$ bits $= 128$ kB, which is too large to fit in the caches, leading to a high number of cache misses.

It is essential to assess the effectiveness of integrating chaos-based encryption with a conventional compression technique for CTE configurations, we next compare with C2. We first perform the compression using \zstd and then encrypt the data using the chaos-based scheme proposed by Wang et al.~\cite{chaos6}. 
We observe that \name results in nearly the same compression ratio as C2~\cite{chaos6}. However, \name is $24\times$ faster than C2~\cite{chaos6}. The reasons for the poor performance of C2~\cite{chaos6} are as follows \circled{1} The algorithm is optimized for images (like most of the chaos-based encryption algorithms).
In order to reduce the performance overheads, the complex computations (strong encryption) are performed on a specified region, which is identified by the \textit{Region of Interest (ROI)}. For the non-ROI regions, a weaker encryption method is utilized.
However, while working with NNs, we must assume that the ROI represents the entire model, as no information regarding the model can be leaked. This results in very high performance overheads. \circled{2} Due to the involvement of the \textit{sine} function, the generation of the chaotic map takes a considerable amount of time. This map is used in the encryption of both ROI and non-ROI regions. \circled{3} The third reason is that the encryption strategy is quite complex and requires a lot of sequential stages that are not hardware-friendly. As a result, the approach results in very poor performance when applied to NN model parameters.

\begin{table}[!h]
\footnotesize
    \centering
    \begin{tabular}{|p{0.08\textwidth}|p{0.35\textwidth}|}
     \hline
     \rowcolor{blue!10}
     Config. & Description\\
     \hline
     E1    & An ETC scheme where \zstd and AES-CBC is used \\
     C1    & A CTE scheme where \zstd and AES-CBC is used \\ 
     C2~\cite{chaos6}    &  A CTE scheme where the encryption scheme is proposed by Wang et al.~\cite{chaos6}. For a fair analysis, We added \zstd for compression\\
     S1{~\cite{helper}}    & The SCE scheme proposed by Ahmad et al.~\cite{helper} \\
     \name  & The proposed SCE implementation \\
          \hline
    \textit{Sec\_deflate} & \name encryption + deflate compression \\
    \textit{Sec\_gzip} & \name encryption + \textit{gzip} compression \\
    \textit{Sec\_bzip} & \name encryption + \textit{bzip} compression \\
    \textit{Sec\_lzma} & \name encryption + \textit{lzma} compression \\
    \hline
    \end{tabular}
    \caption{The different design configurations implemented in the paper}
    \label{tab:configurations}
\end{table}

\subsubsection{A fair comparison between CTE, ETC and SecOComp Schemes}
We implemented our scheme \name along with the baseline CTE (C1) and ETC (E1) schemes for performing a fair comparative analysis between the three schemes. We used Intel's AVX (SIMD 256-bit~\cite{simd}) instructions for further reducing the execution time. We selected the AES-CBC~\cite{aes}) openSSL implementation for the ETC and CTE implementations as it is the most popular and secure cipher according to the NIST standards, and we selected the \textit{zstd} compression scheme for a fair analysis. We present the results in Figures \ref{fig:comp1} and \ref{fig:comp2}. 

We observed that the ETC scheme results in a very poor compression ratio, while CTE results in a marginally increased compression ratio (+1.83\%). Note that although we have utilized the same compression algorithm in both the \name and CTE schemes, the weights in our scheme were permutated before compression. This leads to a 1.83\% decrease in the compression ratio, which is minimal. 
Additionally, our scheme is $82 \%$ faster than ETC and $30.4 \%$ faster than CTE. 


It is clear that ETC schemes are ineffective. As a result, we shifted our focus to different compression schemes using \name (of the form $Sec\_XYZ$). 

\subsubsection{A comparison with other \textit{SecOComp} configurations}
We study the impact of modifying the type of compression scheme on the compression ratio and the execution time of \name (see Figures \ref{fig:comp3} and \ref{fig:comp4}). We observe that \name is $8.62\times$, $8.11\times$, and $8.92\times$ faster than \textit{Sec\_deflate}, \textit{Sec\_gzip}, and \textit{\_lzma}, respectively. \name leads to only a $3.6 \%$ reduction in the compression ratio as compared to \textit{Sec\_lzma} with a  $8.92 \times$ speedup. \textit{Sec\_bzip} leads to the best compression ratio but the scheme is $ 4.99 \times$ slower than \name. The results can easily be correlated with our characterization results (see Section~\ref{sec:char}).

\begin{figure*}[!t]
\subfloat[]
{
    \includegraphics[scale=0.27]{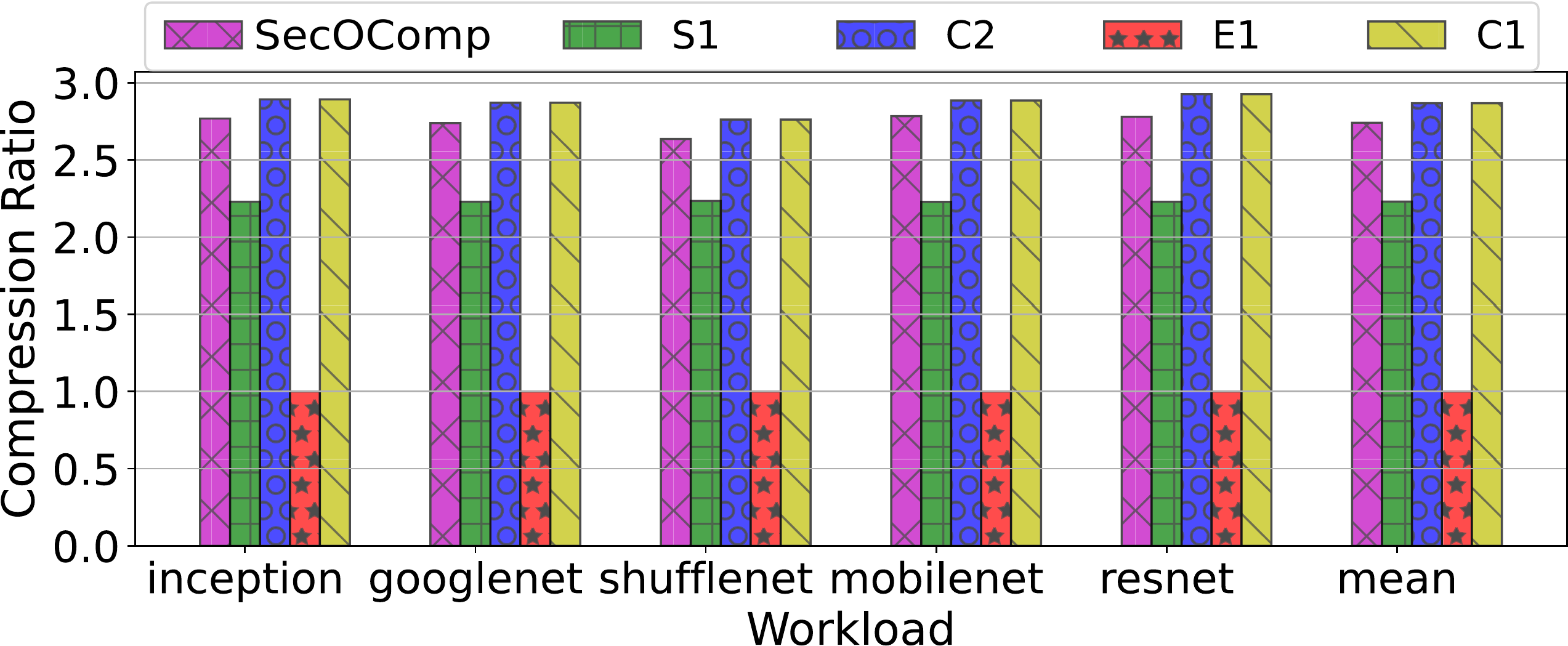}
    \label{fig:comp1}
}
\hfil
\subfloat[]
 { 
    \includegraphics[scale=0.28]{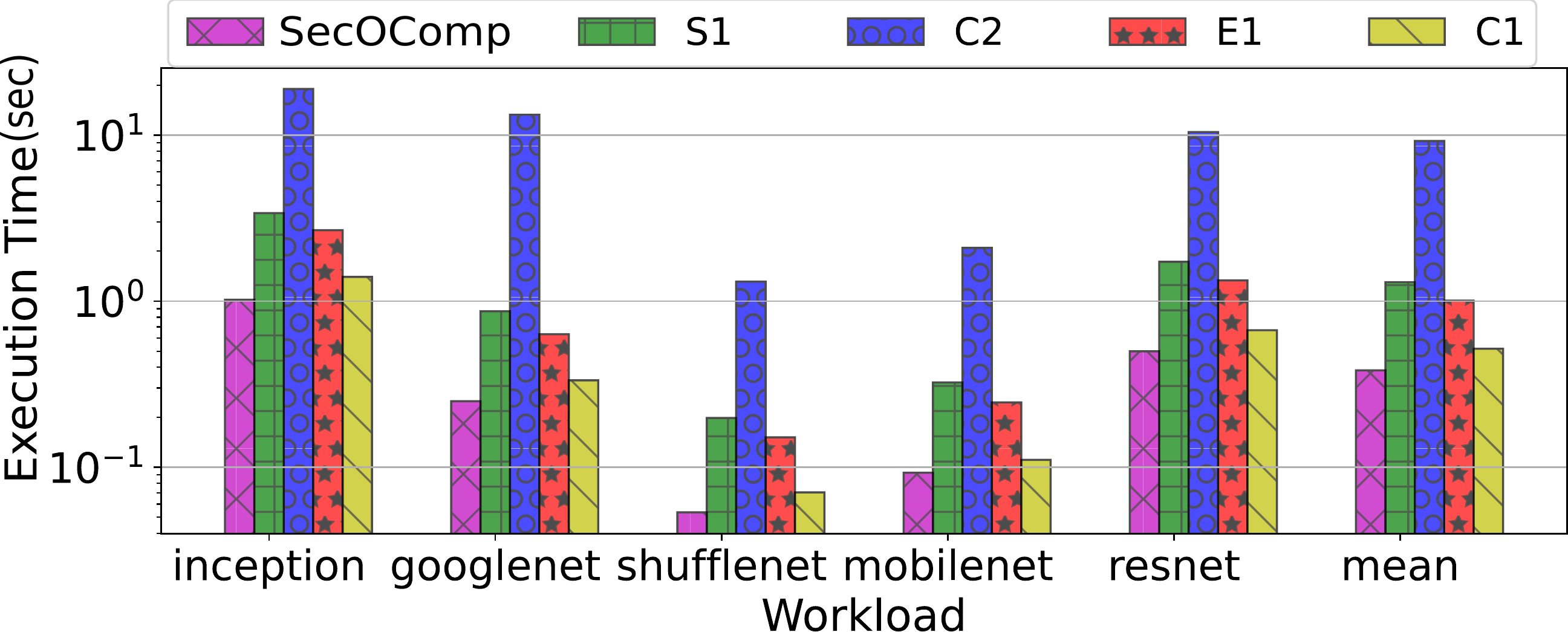}
    \label{fig:comp2}
}

\subfloat[]
{
    \includegraphics[scale=0.28]{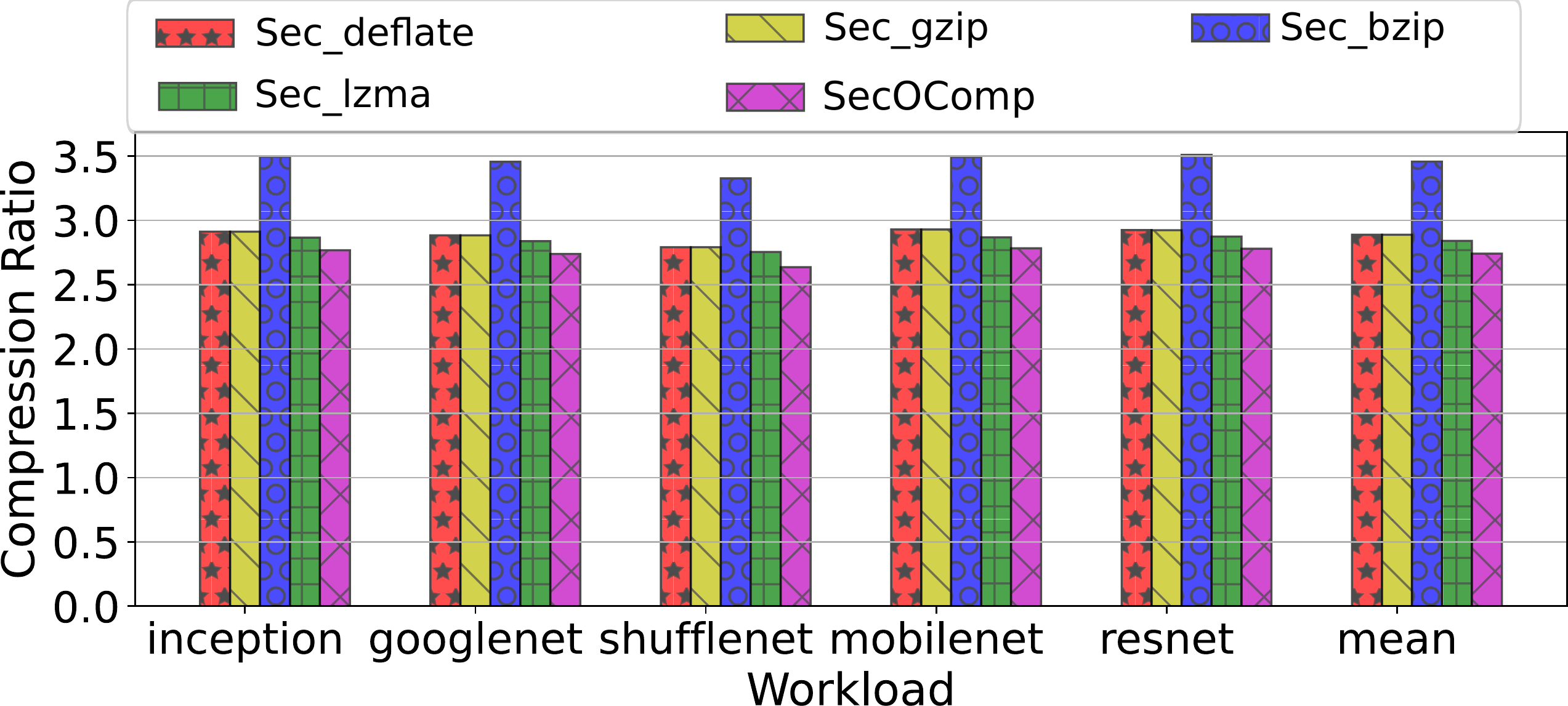}
    \label{fig:comp3}
}
\hfil
\subfloat[]
{
    \includegraphics[scale=0.27]{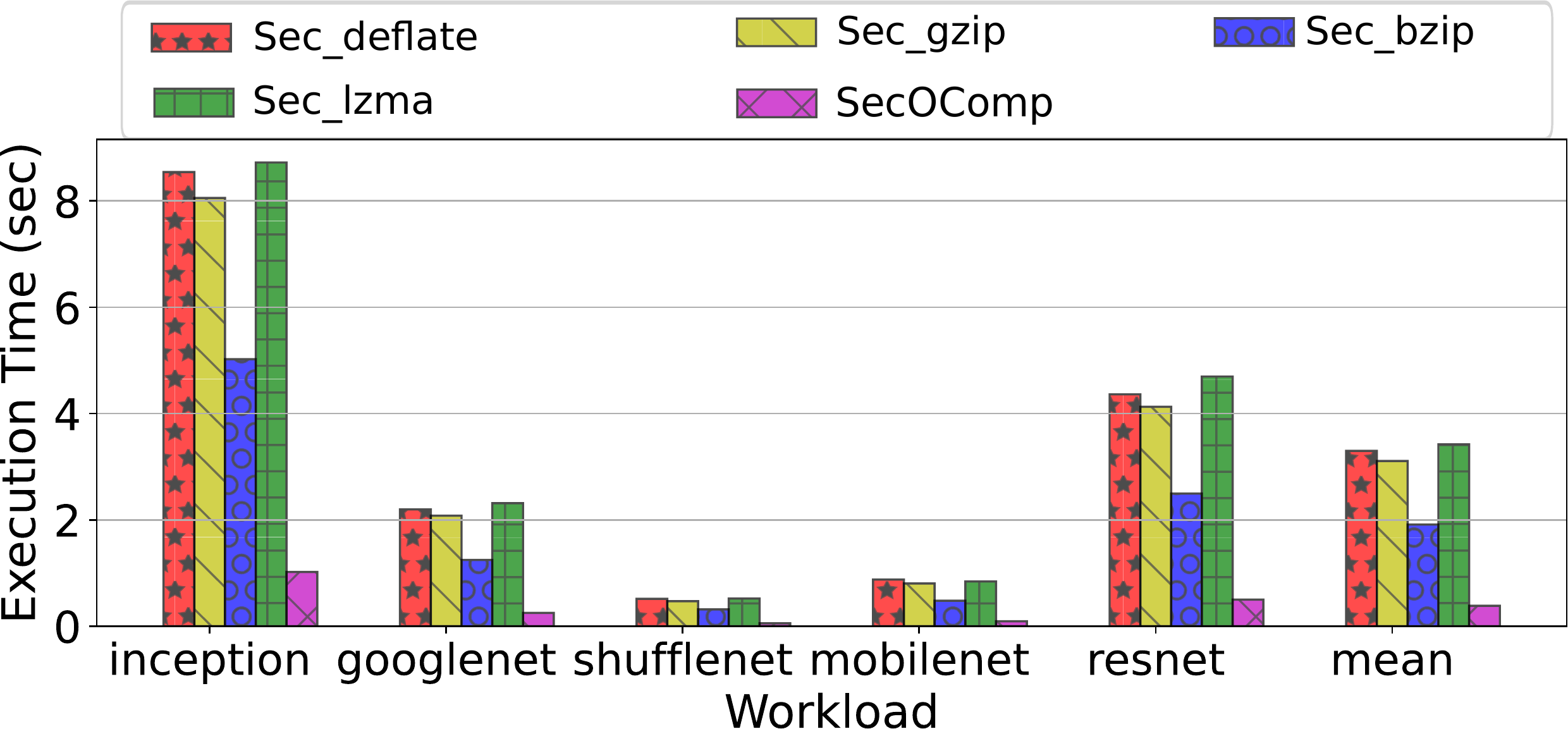}
    \label{fig:comp4}
} 
\caption{A comparison (software-based) on the basis of the \textbf{(a)} compression ratio for various configurations. \textbf{(b)} execution time of various configurations \textbf{(c)} compression ratio for \name configurations \textbf{(d)} execution time of \name configurations}
\end{figure*}

\subsection{Hardware Implementations of \textit{SecOComp} and CTE}
We selected the two best schemes (\name and C1) in terms of the execution time and the compression ratio, respectively, and implemented them in hardware.
The code for \name and CTE(C1) was written in VHDL 2008. We synthesized, placed, and routed the designs using the Cadence Genus tool at the 28 nm ASIC technology node. The compression engine operates at a frequency of 300 MHz, while the permutation and substitution engines operate at 1 GHz. Table \ref{tab:asic} presents the area and power results for each module. Figure \ref{fig:hwcomp} shows a comparison of the execution time in HW. We observe a $23.5\%$ improvement in the execution time (for \name), $19.8\%$ area improvement as well as a $26.11\%$ improvement in power consumption as compared to C1.

\begin{table}[h]
\footnotesize
    \centering
    \begin{tabular}{|c|c|c|}
    \hline
    \rowcolor{blue!10}
    \textbf{Module} & \textbf{Area($\mu m^2$ $\times$ $10^3$)} & \textbf{Power(mW)}  \\
    \hline
    \rowcolor{gray!20}
    \multicolumn{3}{|c|}{\name}\\
    \hline
    Permutation & 0.89  & 0.06\\
    Compression (zstd) & 112.91   & 10.83 \\
    Substitution & 2.26  & 0.17\\
   \textbf{Total} & \textbf{116.06 } &\textbf{11.06} \\
    \hline
    \hline
    \rowcolor{gray!20}
     \multicolumn{3}{|c|}{C1}\\
     \hline
     Compression (zstd) & 112.91   & 10.83 \\
     AES-128 & 31.9 & 4.14 \\
       \textbf{Total} & \textbf{144.81} & \textbf{14.97} \\
    \hline
    \hline
    \rowcolor{gray!20}
    \textbf{Tool} & \multicolumn{2}{c|}{Cadence RTL Compiler, 28 nm} \\
    \hline
    \end{tabular}
    \caption{ASIC Resource and Power Utilization for Various Components for \name and C1}
    \label{tab:asic}
\end{table}

\begin{figure}[!h]
    \centering
    \includegraphics[scale=0.30]{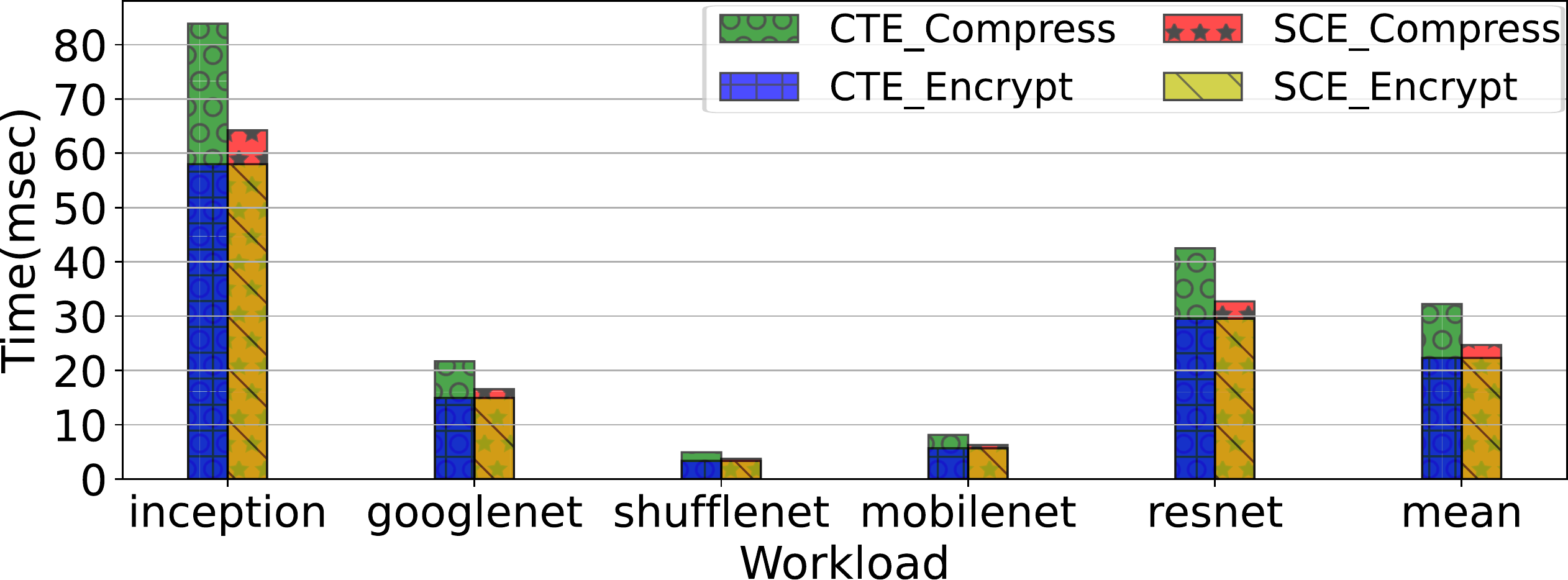}
    \caption{A comparison (HW implementation) of the execution time for \name and C1}
    \label{fig:hwcomp}
\end{figure}

\begin{table*}[h]
\footnotesize
    \centering
    \begin{tabular}{|l|l|l|l|l|l|l|l|l|}
    \hline
    \rowcolor{blue!10}
    & \multicolumn{8}{c|}{\textbf{Key sensitivity} ($\times 10^{-3}$)} \\
   \cline{2-9}
   \rowcolor{blue!10}
   \textbf{Scheme}  & \multicolumn{4}{c}{\textbf{CSI}} & \multicolumn{4}{|c|}{\textbf{Pearson Coefficient}}  \\
    \cline{2-9}
    \rowcolor{blue!10}
    & \textbf{\name} & \textbf{S1} & \textbf{C2} & \textbf{AES} & \textbf{\name} & \textbf{S1} & \textbf{C2} &   \textbf{AES}   \\
     \rowcolor{blue!10}
    &&  ($\times 10^{3}$)  &  &  &  & ($\times 10^{3}$)  & & \\
         \hline
      Inception & 0.256 & 0.076 & 0.579 & 0.074 & 0.260 & 0.076 &  0.124 & 0.028 \\
      
      Googlenet & 0.033 & 0.072 & 0.213 & 0.317 & 0.041 & 0.072 & 0.171 & 0.274 \\
      
      Shufflenet & 0.212 & 0.069 & 0.237 & 0.510 & 0.220 & 0.069 & 0.195 & 0.458 \\
      
      MobileNet & 0.426 & 0.061 & -0.102 & -0.181  & 0.413 & 0.061 & -0.140 & -0.222 \\
      
      Squeezenet & 0.027 & 0.078 & -0.330 &  0.018 & 0.033 & 0.078 & 0.290 & 0.027 \\
      \hline
      \rowcolor{black!8}
      \textbf{Average} & 0.198 & 0.067 &  0.291 & 0.220 & 0.193 & 0.071 &0.184 & 0.201\\
      \hline
    \end{tabular}
    
\vspace{2mm}
    \centering
    \begin{tabular}{|l|l|l|l|l|l|l|l|l|}
    \hline
    \rowcolor{blue!10}
    & \multicolumn{8}{c|}{\textbf{Plaintext sensitivity} ($\times 10^{-3}$)} \\
   \cline{2-9}
   \rowcolor{blue!10}
   \textbf{Dataset}  & \multicolumn{4}{c}{\textbf{CSI}} & \multicolumn{4}{|c|}{\textbf{Pearson Coefficient}} \\
    \cline{2-9}
    \rowcolor{blue!10}
    & \textbf{\name} & \textbf{S1} & \textbf{C2} &  \textbf{AES} & \textbf{\name} & \textbf{S1} & \textbf{C2} & \textbf{AES}  \\
     \rowcolor{blue!10}
    &  & & ($\times 10^{3}$) & & & & ($\times 10^{3}$) &  \\
         \hline
      Inception & 0.616 & -0.065 & 0.124 & 0.171 & 0.614 & -0.094 & 0.124 & 0.125\\
      
      Googlenet  & 0.286 & -0.384 & 0.043 & 0.413 & 0.285 &  -0.416 & 0.043 & 0.369\\
      
      Shufflenet & 0.609 & -2.520 & 0.125 & 0.552 & 0.611 &  -2.528 & 0.125 & 0.615\\
      
      MobileNet & 0.289 & 1.420 & 0.043 & -0.452 & 0.249 &  1.372 &  0.043 & -0.895\\
      
      Squeezenet & 0.443 & -2.901 & 0.125 & 0.287 & 0.441 & -2.904 & 0.124 & 0.333\\
      \hline
      \rowcolor{black!8}
      \textbf{Average} & 0.442 & 1.456 & 0.092 & 0.390 & 0.438 &  1.458 & 0.091 & 0.479 \\
      \hline
    \end{tabular}
    \caption{Key and Plaintext sensitivity analysis  }
    \label{tab:avalanche}
\end{table*}

\section{Security Analysis }
\label{sec:security}
The security of a system can be demonstrated using one of four basic approaches \circled{1} statistical testing (e.g. NIST tests), \circled{2} resistance to known attacks, \circled{3} proof by reduction, and \circled{4} disclosing the scheme to the public, and waiting for an adversary to mount an attack (if possible).
A common approach for establishing the security of any algorithm is by proof by reduction.
We rely on the fact that a standard problem $A$ is extremely difficult to solve. 
We establish the fact that the problem $A$ is reducible to $B$.
If solving $A$ is difficult, then solving $B$ must be equally or more difficult as well. Sadly, there is no known standard solution in this domain for chaos-based encryption. This makes it hard for us to apply proof by reduction.
The remaining strategies are discussed in detail in the next subsections. Our analysis subsumes analyses performed in prior work, especially recent work done on chaos-based encryption in the HW community~\cite{chaos1,chaos2,chaos3,chaos4,chaos5,chaos6,chaos7,chaos8,chaos9,chaos10,chaos11,chaos12,helper}.


\subsection{NIST Tests}
We performed all the tests specified in the NIST suite as shown in Table~\ref{tab:nistTest}.
A test is successful if the $p$-value is greater than a threshold $\alpha$ (set to 0.01 on the lines of prior work~\cite{nist}). We used 100 chunks of data (samples) of a length of 1 million each (for \textit{inception} and \textit{squeezenet}). However, because the model sizes of the other benchmarks is small, we used only 10 samples with a length of 1 million~\cite{nist}. We were unable to achieve the $p$-value for the \textit{random excursions} and \textit{random excursions variant} tests (shown as 0 in the heatmap) as these tests are not applicable because they require a minimum sample length of 1 million and to get a valid $p$-value, we require at least 55 samples~\cite{nist}. Figure~\ref{fig:heatmap} shows the heatmap of $p$-values of all the tests computed using the NIST test suite.
\textit{The pass rate (\# samples passing the test/total \# of samples) of all the tests varies between $0.98-1.0$, which indicates that \name passes {\bf all the 15 NIST tests}. This is as per the documentation provided along with the test suite.}  

We also performed all the NIST tests on the ciphertexts generated by the schemes E1, C1, C2 and S1. We observed that the S1 scheme successfully passed 11/15 NIST tests. 
On the contrary, C2 failed several NIST tests. We also need to note that E1 and C1 passed all the NIST tests as they simply encapsulate AES-CBC.
\begin{figure}[h]
    \centering
    \includegraphics[scale=0.30]{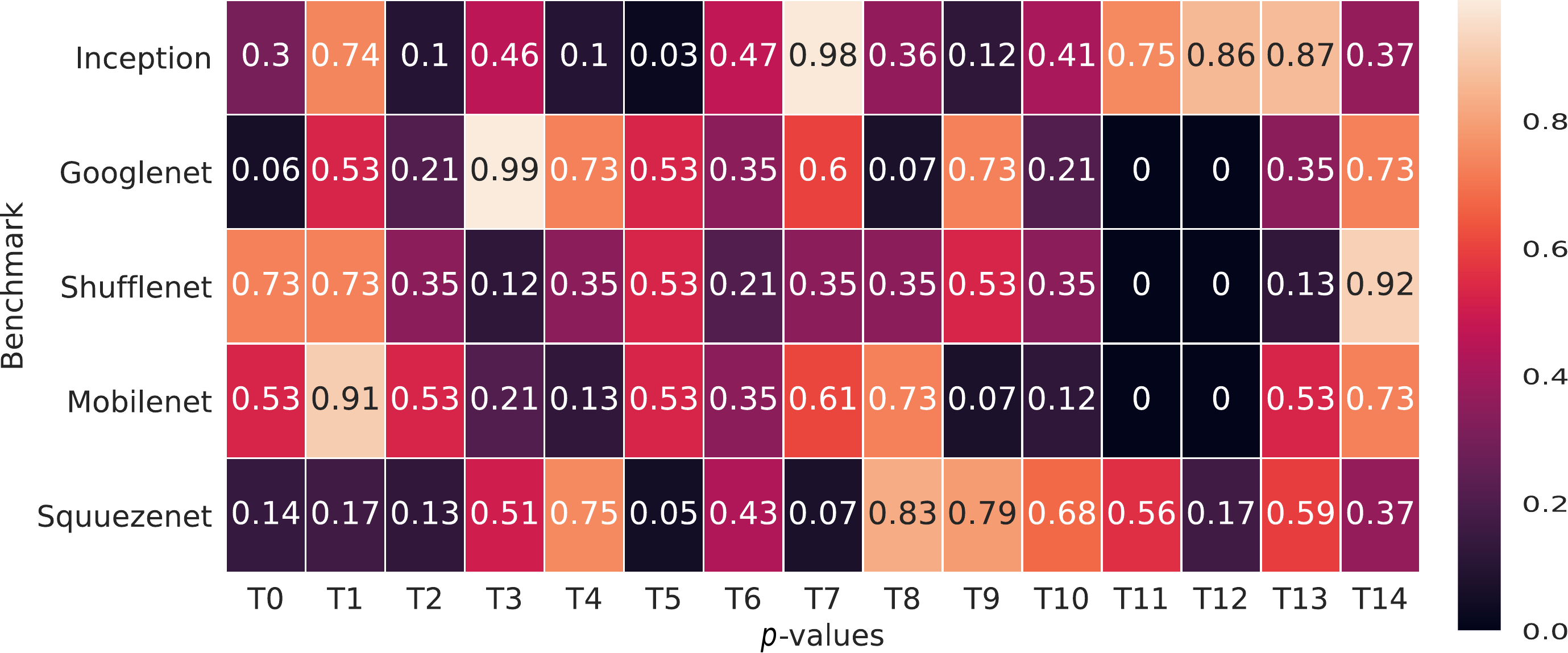}
    \caption{NIST Tests for \name - A heatmap depicting the $p$-values for various benchmarks. A $p$-value$>\alpha$ signifies that the scheme passed the test.
    }
    \label{fig:heatmap}
\end{figure}

\subsection{Correlation Coefficient (CC) Analysis}
 The CC allows us to assess the degree of similarity between the plaintext and the ciphertext. The degree of similarity increases as the correlation value increases. In order to perform a correlation analysis, we calculate the \textit{Pearson correlation coefficients} (using Equation 2) between the plaintexts and the ciphertexts generated using AES-CBC (covers C1 and E1),  S1, C2 and \name. We show the results in Table~\ref{tab:corr}.
An efficient encryption technique will result in a correlation coefficient with a very low absolute value. We observe that \name can provide similar levels of encryption as the AES-CBC cipher.

\begin{table}[h]
\footnotesize
    \centering
    \begin{tabular}{|l|l|l|l|l|}
    \hline
    \rowcolor{blue!10}
    \textbf{Scheme} & \multicolumn{4}{c|}{\textbf{Correlation Coeff. ($ \times 10^{-3}$)}} \\
    \cline{2-5}
    \rowcolor{blue!10}
      & \textbf{AES-CBC} & \textbf{S1} & \textbf{C2} & \textbf{SecOComp } \\
      \hline
      Inception   & 0.017 & 0.144 & 0.199 & 0.151\\
      Googlenet   & 0.161 & 0.554 & -0.140 & 0.039\\
      Shufflenet & 0.191 & -0.317 & 0.777 & 0.025\\
      MobileNet & 0.215 & -0.607 & 0.759 & 0.285\\
      Squeezenet & 0.161 & 0.380 & -0.188 & 0.269\\
      \hline
      \rowcolor{black!8}
      \textbf{Average} & 0.151 & 0.400 & 0.425 & 0.153 \\
      \hline
    \end{tabular}
    \caption{Correlation Between the Plaintext and Ciphertext in AES-CBC, S1, C2 and \name}
    \label{tab:corr}
\end{table}

\subsection{Key Space Analysis}
The Kerckhoff’s principle states that the security of an algorithm should be dependent only on the secret key. 
An efficient encryption algorithm should have a large keyspace that is resistant to brute-force attacks~\cite{helper,chaos6,securityana}. In general, the keyspace is the product of the number of possible subkeys used in each phase of the algorithm. $K_S= \prod K_i$, where $K_S$ is the keyspace and $K_i$ is the number of subkeys in each phase of the algorithm.
Our proposed technique is based on a permutation-and-substitution procedure involving three chaotic maps. Thus, our key consists of three components: the permutation key $K_p$ and the XOR engine keys $K_{s1}$ and $K_{s2}$.
$K_p$ is composed of one initial condition (corresponding to the \textit{Logistic} map), one secret parameter ($\mu$) and a secret threshold value. Key $K_{s1}$ consists of two initial conditions (corresponding to the \textit{Henon} map) and two secret parameters ($a$, $b$), whereas key $K_{s2}$ consists of three initial conditions (corresponding to the \textit{Lorenz} map) and three secret parameters ($\sigma$, $\rho$, $\beta$).
The computational precision of 32-bit fixed point number is approximately $10^8$. 

The total keyspace of the proposed scheme can be written as (approximately): $K_S= K_{s1}\times K_p \times K_{s2}=10^{8\times7 + 6\times8} =10^{104}$.

\subsection{Sensitivity Analysis}
We perform a sensitivity analysis to examine the impact of a little modification in the plaintext or the secret key on the entire ciphertext (referred to as {\em diffusion} and {\em confusion}, respectively). In line with the prior work, we use the \textit{Pearson's correlation coefficient (CC)} and the \textit{Cosine Similarity Index (CSI)}. CSI is a metric that is used to measure the similarity between two vectors. It is computed as $  \frac{X.Y}{|X| \times |Y|}$.
The value of CSI lies between -1 and 1.
The lower the magnitude of the value, the lower the correlation between the two datasets. 
We use the in-built function from the \textit{sklearn 0.24.1} library~\cite{sklearn} to compute the CSI.

\subsubsection{Key Sensitivity Analysis }
The key sensitivity analysis determines the effect of a 1-bit change in the key on the ciphertext.
We use the CSI and CC to compare the similarity of ciphertexts generated by changing a single bit of the key as shown in Table~\ref{tab:avalanche}. We compare the results with the most recent SCE implementation S1, C2 and AES-CBC (covers C1 and E1). We observe that our values are nearly the same as the AES-CBC.

\subsubsection{Plaintext Sensitivity (Avalanche Effect) Analysis }
Similar to the key sensitivity analysis, plaintext sensitivity indicates the effect of a single-bit change in the plaintext on the ciphertext. An encryption method must produce fully distinct ciphertexts from a group of plaintexts containing small changes. We examine the similarity between the ciphertexts generated by altering a single bit of the plaintext using the CSI and CC as shown in Table~\ref{tab:avalanche}. We observe similar results as the earlier experiment.

\subsection{Classical Attacks}
There are some well-known classical attacks in the crypt-analysis literature, such as the \circled{1} \textit{Known Plaintext Attack} where an attacker \textit{knows} plaintexts and the corresponding ciphertexts.
     \circled{2} \textit{Chosen Plaintext Attack} where an attacker can select a plaintext to pass through the encryption algorithm and obtain its corresponding ciphertext.
     \circled{3} \textit{Chosen Ciphertext Attack} where an attacker can choose a ciphertext to pass through the decryption algorithm and obtain its corresponding plaintext.
         \circled{4} \textit{Only Ciphertext Attack} where an attacker has an access to ciphertexts.\\
       It has been theoretically proven that out of these attacks, the chosen plaintext attack is the most powerful~\cite{securityana}. Consequently, if a cryptosystem can withstand this attack, it can withstand other types of attacks as well. Several authors~\cite{securityana,chaos6,chaos13,chaos14} mention that if an algorithm has strong confusion and diffusion properties, they can withstand such attacks. The preceding section already demonstrated that \name has significant confusion and diffusion.

\subsection{Crypt-analysis of \textit{SecOComp}}
Chen et al.~\cite{attack3} propose a theoretical crypt-analysis framework for chaos-based encryption systems. This is the most recent and comprehensive work in this domain. The authors perform cryptanalysis on the algorithms, which satisfy the following property and outline methods to break them:  $C_1 \oplus C_2 = Perm_K(P_1 \oplus P_2)$, where $C_i=Perm_K(P_i) \oplus K_1$. $(C_1,C_2)$ and $(P_1, P_2)$ are a pair of ciphertexts and the corresponding plaintexts, respectively. $\oplus$ denotes a XOR operation, $K_1$ represents the secret key and $Perm_K$ denotes a permutation operation dependent on the secret key $K$. Fortunately, \name does not satisfy the above-mentioned property because the ciphertext is a complex function of the compressed permuted data and the secret key.
This statement of ours has been theoretically and experimentally validated. 

\subsection{Additional Results}

We examine the efficiency and security of our scheme when applied to different types of datasets utilizing the identical system configuration described in Table \ref{tab:config}.
\subsubsection{Datasets Description}
We demonstrate the effectiveness and security of our technique for various images.
We investigated five standard images from the USC-SIPI Image Database\cite{img}: \textit{baboon}, \textit{jelly}, \textit{peppers}, \textit{tank}, and \textit{woman}.
We also take into account other datasets, which are obtained from the official Kaggle website.
The details of the datasets are present in Table \ref{tab:datasets}. Both the USC-SIPI Image dataset and the Kaggle dataset have received considerable attention from the scientific community and have been extensively used~\cite{kaggle1,image1,image2}.  

\begin{table}[h]
\footnotesize
    \centering
    \begin{tabular}{|p{4.3cm}|p{1.21cm}|p{0.73cm}|p{0.8cm}|}
    \hline
    \textbf{Dataset Description} & \textbf{Category}   & \textbf{Size} & \textbf{Notation}  \\
    \hline
     Covid cases and deaths worldwide~\cite{D1} &  Healthcare & 8 kB & D1\\
     \hline
     Top 1000 movies by IMDB rating~\cite{D2} & User study & 53 kB & D2 \\
    \hline
    $CO_2$ emission by country~\cite{D3} & Weather &117 kB & D3 \\
    \hline
     Data of 1000+ Amazon product's rating and reviews~\cite{D4} & User study & 2 MB & D4 \\
     \hline
    3000 conversations dataset for chatbots~\cite{D5} & Phone & 69 kB & D5 \\
    \hline
    Medical cost personnel datasets - Insurance forecast by using linear regression~\cite{D6} & User study & 16 kB & D6 \\
    \hline
    Numerically generated ECG signal~\cite{D7} & Healthcare & 1 MB & D7\\
    \hline
    \end{tabular}
    \caption{Description of the datasets}
    \label{tab:datasets}
\end{table}
\vspace{-2mm}
\subsubsection{Performance and Security Analysis}
Figures \ref{fig:comp5} and \ref{fig:comp6} present the execution time and compression ratio for all the selected images.
Figures \ref{fig:comp7} and \ref{fig:comp8} demonstrate the execution time and compression ratio for other datasets.  We observe that \name is able to provide an optimal ratio between the execution time and the compression ratio. We performed an extensive security analysis and we present the results in Tables ~\ref{tab:corr_dataset} and ~\ref{tab:security_dataset}. We also conducted NIST tests to evaluate the algorithm's security.
We see that \name passed all necessary NIST tests and can give adequate security guarantees.

\begin{table*}[h]
  \footnotesize
    \centering
    \begin{tabular}{|l|l|l|l|l|l|l|l|l|}
    \hline
    \rowcolor{blue!10}
    & \multicolumn{8}{c|}{\textbf{Key sensitivity} ($\times 10^{-2}$) } \\
   \cline{2-9}
   \rowcolor{blue!10}
   \textbf{Dataset}  & \multicolumn{4}{c}{\textbf{CSI}} & \multicolumn{4}{|c|}{\textbf{Pearson Coefficient}}  \\
    \cline{2-9}
    \rowcolor{blue!10}
    & \textbf{\name} & \textbf{S1} & \textbf{C2} & \textbf{AES} & \textbf{\name} & \textbf{S1} & \textbf{C2} &   \textbf{AES}   \\
         \hline
      D1 & -0.1010 & 0.5844 & 0.0393 & 0.7483 & -0.1037 & 0.5805 &  0.0349 & -0.0683 \\
      
      D2 & 0.0135 & 2.0347 & 0.5224  & 0.7477 &  -4.9005 & 2.0405 & 0.5243   &  -1.2372\\
      
      D3 & -0.0416 & 7.6540 & 0.0771 & 0.7484 & -0.0411 & 7.6487 &  0.0705 & -0.0381 \\
      
      D4 & 0.0212 & 6.4575 & 0.1252  & 0.7491 & 0.0172 & 6.4553 & 0.1222  & 0.1986 \\
      
      D5 & 0.0692 & 1.1027 & -0.0305 & 0.7485 & 0.0662 &  1.0100 & -0.0351  & 0.0663 \\

      D6 & -0.2282 & 1.0739 & -0.1131 & 0.7489 & -0.2351 &  1.0735 & -0.1152  &  -0.0202\\

      D7 & 0.1219 & 9.5453 & -1.2454 & 0.7489  & 0.1189  &  9.5470 & -1.2559  &  0.1035 \\
      \hline
      \rowcolor{blue!10}
      \multicolumn{9}{|c|}{\textbf{Images}}\\
      \hline
      baboon & 0.1884 & 8.2555 & -0.0634 & 0.7481 & 0.1831 & 8.2542  &  -0.0693 & -0.1135 \\
      
      jelly & 0.1149 & 6.3944 & 0.0672 & 0.7479 & 0.1081 & 6.3907 &  0.0635 & -0.0318 \\
      
      peppers & 0.1096 & 6.4588 & -0.0526 & 0.7485 & 0.1049 &  6.4563 &  -0.0584 & -0.0523 \\
      
      tank & 0.2192  & 6.5114 &  0.2410 & 0.7484  & 0.2134 & 6.5075 & 0.2356  & -0.0569 \\
      
      woman & 0.2126 & 0.3014 & 0.0984 & 0.7486 & 0.2114 & 0.2948 & 0.0937  & 0.0491 \\
      \hline
    \end{tabular}
    
\vspace{2mm}
    \centering
    \begin{tabular}{|l|l|l|l|l|l|l|l|l|}
    \hline
    \rowcolor{blue!10}
    & \multicolumn{8}{c|}{\textbf{Plaintext sensitivity} } \\
   \cline{2-9}
   \rowcolor{blue!10}
   \textbf{Dataset}  & \multicolumn{4}{c}{\textbf{CSI}} & \multicolumn{4}{|c|}{\textbf{Pearson Coefficient}} \\
    \cline{2-9}
    \rowcolor{blue!10}
    & \textbf{\name} & \textbf{S1} & \textbf{C2} &  \textbf{AES} & \textbf{\name} & \textbf{S1} & \textbf{C2} & \textbf{AES}  \\
         \hline
      D1 & 0.8385 & -0.0515 & 0.0436 & 0.7486 & 0.8385 & -0.0515 & 0.0436  &  0.0003\\
      
      D2 & 0.0063 & -0.0151 & 0.0460 & 0.7521 & 0.0061  &  -0.0151 &  0.0460 & 0.0019  \\
      
      D3 & -0.0074  & -0.0553 & 0.0432 & 0.7479 & -0.0074 & -0.0553 &  0.0432 & -0.0009 \\
      
      D4 & 0.0039 & -0.0520 & 0.0429 & 0.7485  & 0.0039 & -0.0521 &  0.0429 &  0.0001\\
      
      D5 & 0.0002 & 0.0774 & 0.1235 & 0.7485 & 0.0002 &  0.0774 & 0.1235  & 0.0002\\

      D6 & 0.0014 & -0.0095  & 0.1204 & 0.7491 & 0.0014 &  -0.0095 &  0.1203 & 0.0015 \\

      D7 & -0.0089 & -0.0223  & 0.0457 & 0.7503 & -0.0091 & -0.0223  & 0.0456  & 0.0002 \\
      \hline
      \rowcolor{blue!10}
\multicolumn{9}{|c|}{\textbf{Images}}\\
      \hline
    baboon & 0.0024 & -0.0501 & 0.0433 & 0.7489 & 0.0024 &  -0.0501 & 0.0433  & 0.0016 \\
      
      jelly & -0.1538& -0.0233  & 0.1188 & 0.7469 & -0.1539 & -0.0233 & 0.1188  &  -0.0013\\
      
      peppers & 0.1254 & -0.0208 & 0.0999 & 0.7487 & 0.1253 & -0.0208 &  0.098  & 0.0011\\
      
      tank & 0.2912  & -0.0217 & 0.0433 & 0.7484 & 0.2912  & -0.0218 & 0.0433  &  -0.0011\\
      
      woman & 0.1029 & -0.0039 & 0.1190 & 0.7473 & 0.1029 &  -0.0040 & 0.1189  & -0.0032 \\
      \hline
    \end{tabular}
    \caption{Key and Plaintext sensitivity analysis for different datasets}
    \label{tab:security_dataset}
\end{table*}

\begin{table}[h]
    \footnotesize
    \centering
    \begin{tabular}{|l|l|l|l|l|}
    \hline
    \rowcolor{blue!10}
    \textbf{Dataset} & \multicolumn{4}{c|}{\textbf{Correlation Coeff. ($\times 10^{-2}$) }} \\
    \cline{2-5}
    \rowcolor{blue!10}
      & \textbf{AES-CBC} & \textbf{S1} & \textbf{C2} & \textbf{SecOComp } \\
      \hline
      D1 & -0.396 & 3.141 & -1.351 & -0.248\\
      D2 & 0.398 & 0.377 & -0.129 & 0.092\\
      D3 & -0.017 & 0.970 &  0.137 & 0.219\\
      D4 & -0.020 & 0.226 & 0.070 & -0.325\\
      D5 & -0.068 & 0.262 & 0.156  & 0.011\\
      D6 & 0.128 & -0.618 & 0.198 & -0.097\\
      D7 & -0.188 & -0.618 &  0.091 & -0.021\\
      \hline
      \rowcolor{blue!10}
      \multicolumn{5}{|c|}{\textbf{Images}}\\
      \hline
      baboon & 0.106 & 3.394 & -0.085 & -0.868\\
      jelly & -0.085 & 7.007 & -0.125 & 1.878\\
      peppers & 0.006 & -3.916 & 0.248 & -0.177\\
      tank & 0.186 & 1.847 & 0.018 & -1.135\\
      woman & 0.162 & -0.959 & -0.030 & 3.457\\
      \hline
    \end{tabular}
    \caption{Correlation Between the Plaintext and Ciphertext in AES-CBC, S1, C2 and \name}
    \label{tab:corr_dataset}
\end{table}

\begin{figure}[h]
    \centering
    \includegraphics[scale=0.27]{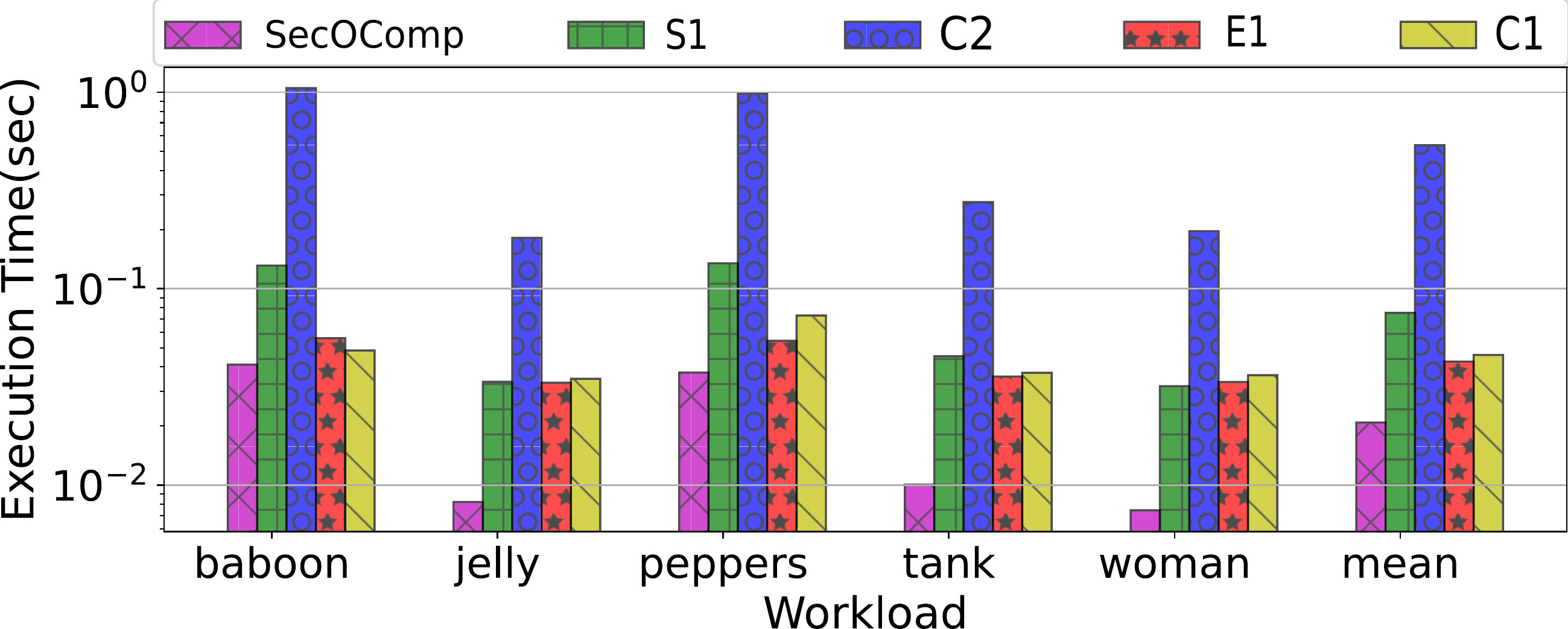}
    \caption{A comparison of the execution time for different images}
    \label{fig:comp5}
\end{figure}
\begin{figure}[h]
    \centering
    \includegraphics[scale=0.27]{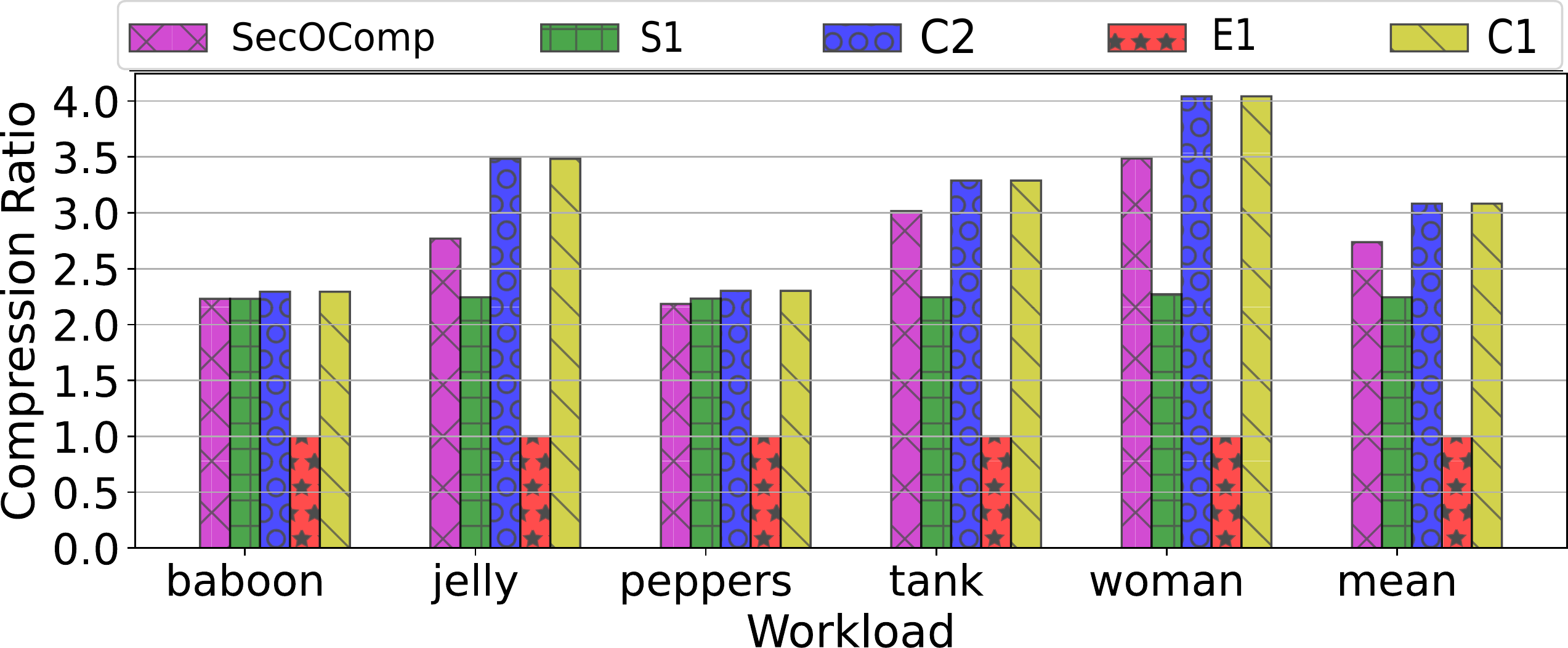}
    \caption{A comparison of the compression ratio for different images}
    \label{fig:comp6}
\end{figure}
\begin{figure}[H]
    \centering
    \includegraphics[scale=0.27]{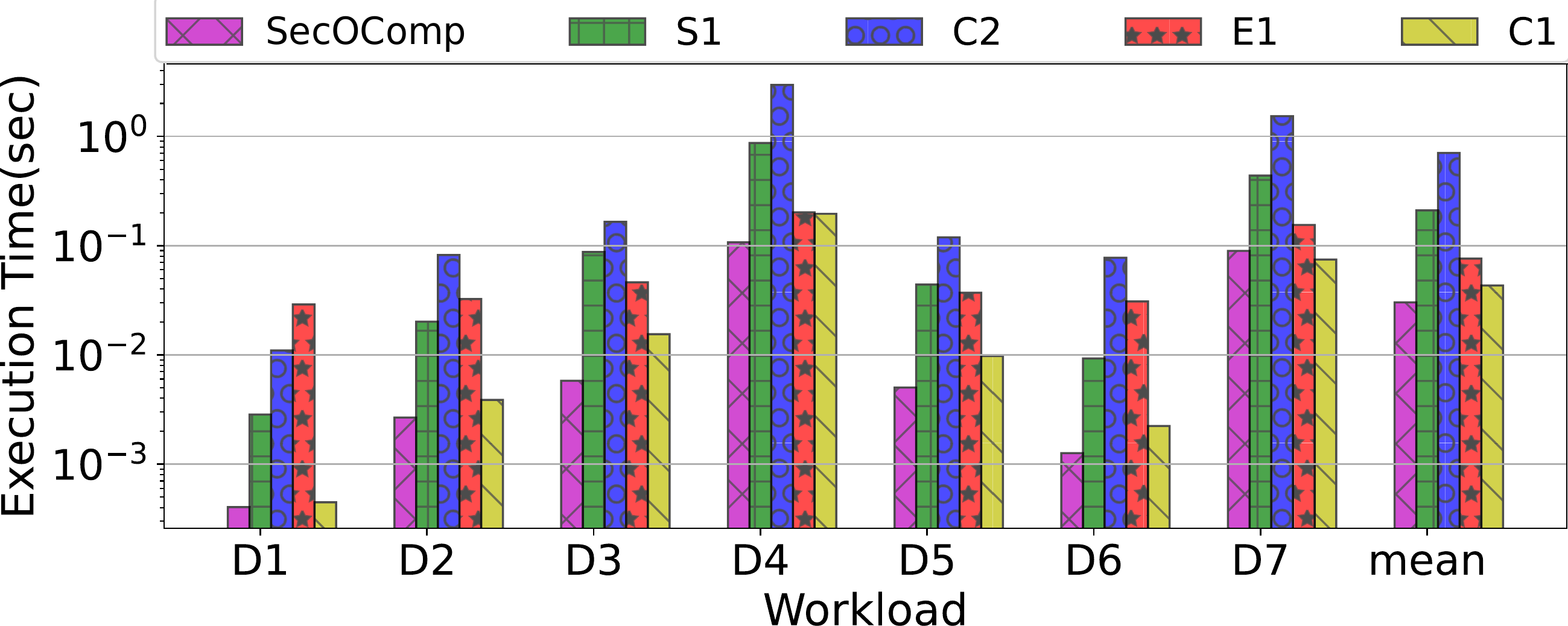}
    \caption{A comparison of the execution time for different datasets}
    \label{fig:comp7}
\end{figure}
\begin{figure}[h]
    \centering
    \includegraphics[scale=0.27]{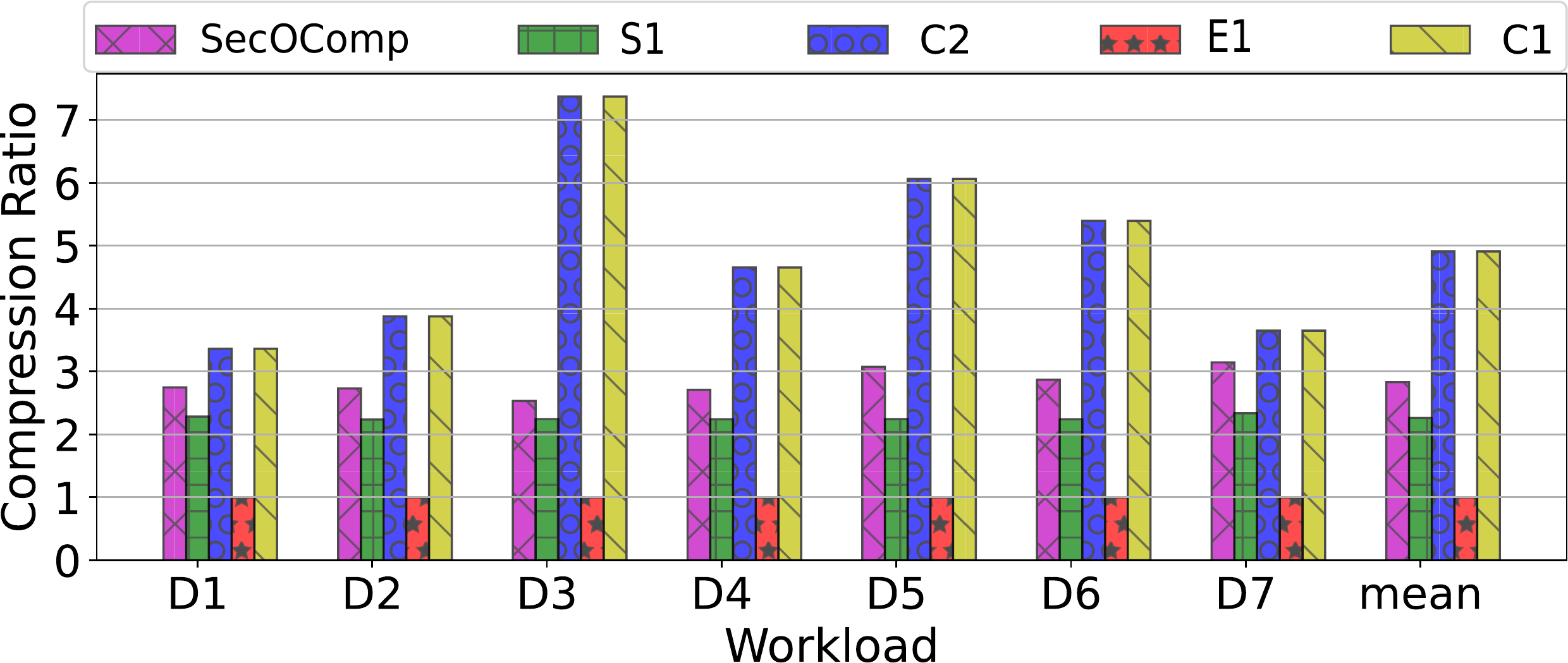}
    \caption{A comparison of the compression ratio for different datasets}
    \label{fig:comp8}
\end{figure}




\section{Related Work}
\label{sec:RW}
SCE schemes have been proposed in the context of images. Wang et al.~\cite{double} propose to simultaneously compress and encrypt an image by dividing the image channels (R, G, B) into blocks and perform compression sensing using chaotic maps. The image is further diffused using chaotic maps. Abuturab et al.~\cite{multiple} propose to
simultaneously compress and encrypt images on similar lines. Four images are decomposed into four sub-bands by using wavelet transforms. They fuse the resulting sub-bands to make a single fused image, which is split into R, G, and B channels, which are subsequently compressed and encrypted. Ahmad et al. \cite{helper} propose using two chaotic maps and the Huffman coding system to simultaneously compress and encrypt the data. First, they propose a chaotic map to permute the data. Then, the adjacent values of the permutated data are represented as a unique solution of the CRT. These values are sent to the Huffman compressor for performing the compression operation. Finally, a \logistic map is used to perform a XOR operation. We showed that such algorithms have issues with performance and security.

Some other works~\cite{chaos6,helper} rely on identifying a region of interest (ROI) containing the majority of the information content.
Wang et al.~\cite{chaos6} proposed separating the image into ROI and non-ROI regions, which would then be encrypted using different chaos-based encryption techniques. They use compute-intensive algorithms for the ROI-based region, which lead to a significant performance overhead. The non-ROI region was encrypted using a simpler scheme. This reduced the overall performance overhead. Sadly, the issue with NN models is that the entire model contains a great deal of intellectual property, which should be secured and thus the entire model is in the ROI. Hence, this approach is unsuitable for NN data.

Due to the large size of NNs, we require not only an efficient compression and encryption engine but also several architectural optimizations to speed up the entire process in hardware. To the best of our knowledge, there is no SCE solution for NN data and there is no work in hardware optimizations to implement the same. Retrofitting SCE solutions proposed for images or intelligently combining different compression and encryption methods did not lead to an effective SCE scheme, hence, a bespoke solution \name was proposed.

\section{Conclusion}
\label{sec:conc}

In this study, we propose \name, an SCE algorithm that is able to successfully \cite{aes} compress and encrypt any type of data. We relied on chaos-based encryption techniques because a combination of different chaotic maps to create encryption algorithms has rejuvenated the area of chaotic cryptography, allowing us to easily realize a high level of security. We went a step further by combining chaotic cryptography with a compression technique. Performing some portion of encryption in the shadow of DRAM reads and compression operations allowed us to speed up the entire process and achieve not only good performance and security but also a decent compression ratio.
We also evaluated state-of-the-art SCE solutions and showed that they are unsuitable for not only NN data but also for other miscellaneous data, and our bespoke solution \name is secure, robust, and fast. 



\bibliographystyle{IEEEtranS}
\bibliography{Ref}
\end{document}